\documentclass[prl,aps,twocolumn,superscriptaddress,floatfix,nofootinbib,10pt]{revtex4-2}

\usepackage[utf8]{inputenc}
\usepackage[T1]{fontenc}
\usepackage{amsmath,amssymb}
\usepackage{bm}
\usepackage{booktabs}%
\usepackage{dcolumn}%
\usepackage{etoolbox}
\usepackage{graphicx}
\usepackage{hyperref}
\usepackage{pdfrender}%
\usepackage[separate-uncertainty,free-standing-units,list-units=single]{siunitx}
\usepackage{tabularx}
\usepackage{xcolor}
\usepackage{xspace}
\usepackage[nameinlink,capitalise]{cleveref}
\newcounter{smref}
\crefname{smref}{Supplemental Material}{Supplemental Material}
\crefformat{smref}{#2SM#3}

\AtBeginEnvironment{thebibliography}{\catcode`\_=12\relax}

\xspaceaddexceptions{+-}
\DeclareRobustCommand{\Og}{\ensuremath{\Omega_\mathrm{g}}\xspace}
\DeclareRobustCommand{\Om}{\ensuremath{\Omega_\mathrm{m}}\xspace}
\DeclareRobustCommand{\cs}{\ensuremath{c_\mathrm{s}}\xspace}
\DeclareRobustCommand{\LCDM}{\ensuremath{\Lambda\mathrm{CDM}}\xspace}

\DeclareRobustCommand{\wwCDM}{\ensuremath{w_0w_a\mathrm{CDM}}\xspace}
\DeclareRobustCommand{\Gc}{\ensuremath{G_\mathrm{cosmo}}\xspace}
\DeclareRobustCommand{\GN}{\ensuremath{G_\mathrm{N}}\xspace}
\DeclareRobustCommand{\Planck}{\textit{Planck}\xspace}
\DeclareRobustCommand{\Hillik}{\texttt{Hillik}\xspace}
\DeclareRobustCommand{\SPA}{\texttt{SPA}\xspace}
\DeclareRobustCommand{\lite}{\texttt{lite}\xspace}

\AtBeginDocument{\RenewCommandCopy\qty\SI}
\DeclareSIUnit\parsec{pc}

\providecommand\mnras{MNRAS}
\providecommand\apj{ApJ}
\providecommand\apjl{ApJL}
\providecommand\aap{A\&A}
\providecommand\jcap{JCAP}
\providecommand\prd{Phys. Rev. D}

\begin{document}

\title{Is DESI Seeing Dynamical Dark Energy, or a Cosmic Glitch in Gravity?}

\author{Robin Y. Wen}
\affiliation{California Institute of Technology, Pasadena, CA 91125, USA}

\author{Lukas Tobias Hergt}
\affiliation{Universit\'e Paris-Saclay, CNRS/IN2P3, IJCLab, 91405 Orsay, France}

\author{Niayesh Afshordi}
\affiliation{Waterloo Centre for Astrophysics, University of Waterloo, 200 University Ave W, Waterloo, Ontario N2L 3G1, Canada}
\affiliation{Department of Physics and Astronomy, University of Waterloo, 200 University Ave W, Waterloo, Ontario N2L 3G1, Canada}
\affiliation{Perimeter Institute for Theoretical Physics, 31 Caroline St N, Waterloo, Ontario N2L 2Y5, Canada}

\author{Douglas Scott}
\affiliation{Department of Physics \& Astronomy, University of British Columbia, Vancouver, BC V6T 1Z1, Canada}

\date{\today}

\begin{abstract}
Combined baryon acoustic oscillation and cosmic microwave background data have begun to mildly favor departures from \LCDM. This is usually framed as 2-parameter dynamical dark energy. However, a 1-parameter ``cosmic glitch in gravity'' model, whose cosmological gravitational coupling differs from Newton's constant as in Ho\v{r}ava gravity, fits at least as well. \Planck+ACT+SPT and DESI~DR2 give $\Gc/\GN=0.9920\pm0.0025$, a $3.3\,\sigma$ preference for weaker cosmological gravity, robust to adding CMB lensing and supernovae. Fitted to the CMB alone, the glitch predicts DESI distances; $w_0w_a$CDM does not.
\end{abstract}

\maketitle

The discovery of cosmic acceleration established that gravity and the vacuum can be tested on cosmological horizon scales. The Dark Energy Spectroscopic Instrument (DESI) has now sharpened that test: its baryon acoustic oscillation (BAO) measurements, when combined with cosmic microwave background (CMB) anisotropy and in some cases supernova (SN) data, show a mild but persistent preference for departures from \LCDM~\cite{25DESI_DR2BAO,25Camphuis_SPT3Gd1,25HeroldKarwal_DDE,26Li_DDE_ACTSPTPlanck,26Hergt_BMCtension}. The common language for this departure is the Chevallier--Polarski--Linder (CPL) form ${w(a)=w_0+w_a(1-a)}$~\cite{01Chevallier_CPL,03Linder_CPL}. \wwCDM is an effective description, however, not a theory: it introduces two phenomenological parameters whose physical origin is left open.

Here we ask a simpler question. Are the data really asking for a 2-parameter evolving dark energy model, or could they be responding to a single physical mismatch between gravity on local and cosmological scales? The cosmic-glitch-in-gravity (CGG) model~\cite{24WenCGG,24WenCGG_essay,Robbers:2007ca} parameterizes such a mismatch by
\begin{equation}
        \Og \equiv 1-\frac{\GN}{\Gc},
        \label{eq:Og}
\end{equation}
where \GN is the locally measured Newtonian coupling and \Gc is the coupling entering the Friedmann equation. In a flat universe this background evolution corresponds to an effective dark-energy density
\begin{equation}
        \rho_\mathrm{DE}
        =
        \frac{\Og \rho_\mathrm{nonDE}+\rho_\Lambda}
             {1-\Og},
        \label{eq:rho_cgg}
\end{equation}
where $\rho_\mathrm{nonDE}$ contains matter, radiation, and neutrinos. The \LCDM limit is $\Og=0$. Negative \Og means $\Gc<\GN$, so the homogeneous Universe gravitates slightly more weakly than local systems.

This deformation is minimal in parameter count, but not entirely ad hoc. A difference between the cosmological and Newtonian gravitational couplings is a generic low-energy signature of gravity theories with a preferred foliation, including Ho\v{r}ava--Lifshitz gravity, Einstein-aether theory, and the quadratic cuscuton gravity limit~\cite{09Hovrava,10Jacobson_Aether,07CuscutonCosmo,09CuscutonHovrava,10Pons_minimal_Horava,25JacobsonPulakkat_mHg}. In the low-energy Ho\v{r}ava parameterization, the same mismatch is commonly written
\begin{equation}
        \frac{\GN}{\Gc}
        = 1 - \Og
        = 1 + \frac{3}{2}(\lambda-1),
        \label{eq:horava_mapping}
\end{equation}
so the GR value is $\lambda=1$, while a negative CGG parameter corresponds to $\lambda>1$~\cite{09CuscutonHovrava}.

In generic khronometric/Ho\v{r}ava gravity the cosmological Newton constant is only one combination of a larger set of Lorentz-violating couplings. Big bang nucleosynthesis, which probes the coupling in the Friedmann equation directly, gives $\Gc/\GN=0.99^{+0.06}_{-0.05}$ at $2\,\sigma$~\cite{20Alvey_BBN_G}, in this notation $|\Og|\lesssim0.06$, while existing cosmological analyses of low-energy Ho\v{r}ava gravity quote stronger percent-level limits before imposing preferred-frame constraints~\cite{14Yagi_HoravaPulsars,21Frusciante_HoravaGW}. Solar-System tests constrain the preferred-frame parameterized post-Newtonian~(PPN) parameters to $|\alpha_1|\lesssim10^{-4}$ and $|\alpha_2|\lesssim10^{-7}$, and in the corresponding tuned regions $\Gc/\GN$ can be forced much closer to unity~\cite{14Yagi_HoravaPulsars,18Oost_AetherGW170817,21Frusciante_HoravaGW}. However, in the incompressible (or cuscuton) limit, where the speed of sound tends to infinity ($c_\mathrm{s}\to\infty$), the theory reduces to GR in flat spacetime, and thus only cosmological constraints remain. Put differently, the local tests constrain the khronon propagation speed rather than \Og. In the khronometric couplings $(\alpha,\beta,\lambda)$ of Ref.~\cite{25KovachikSibiryakov_khronoBH}, whose $\lambda$ is our $\lambda-1$, the khronon sound speed obeys $\cs^2=(\beta+\lambda)/\alpha$, observations of the source GW170817 force $|\beta|\lesssim10^{-15}$, and for $\alpha\ll\lambda$ the preferred-frame parameter reduces to $\alpha_2\simeq-\alpha/2$, so that $|\alpha_2|\lesssim10^{-7}$ gives $\alpha\lesssim2\times10^{-7}$ and hence
\begin{equation}
    \cs^2 \gtrsim \frac{\lambda-1}{2\times10^{-7}} \simeq 3 \times 10^4 \left(\frac{\lambda-1}{0.006}\right).
    \label{eq:cs_bound}
\end{equation}
Other probes constrain directions largely orthogonal to the background ratio. The joint GW170817/GRB~170817A observation bounds the tensor propagation speed at the $10^{-15}$ level, which in Einstein-aether notation implies a very small tensor-speed coupling, and binary-pulsar timing further constrains strong-field preferred-frame effects and dipole radiation~\cite{17Abbott_GWGRBspeed,18Oost_AetherGW170817,14Yagi_HoravaPulsars,21Gupta_AetherPulsars}. For black holes, regular slowly moving solutions appear only in special regions that overlap those selected by GW170817 and Solar-System tests~\cite{19Ramos_HoravaBH}. In summary, the only empirically viable low-energy Ho\v{r}ava--Lifshitz models have a nearly incompressible khronon (the scalar degree of freedom), luminal gravitons, and are detectable only via a cosmic glitch in the gravitational coupling, \Og.

In this Letter, we combine BAO data from DESI~DR2~\cite{25DESI_DR2BAO} with the \Hillik likelihood~\cite{26Tristram_Hillik} for the primary CMB, which combines \Planck-PR4~\cite{20Npipe,24Tristram_PR4} at large angular scales ($\ell^{TT}\!<\!1825$, $\ell^{TE}\!<\!1075$, $\ell^{EE}\!<\!825$), with the Atacama Cosmology Telescope~(ACT)~DR6~\cite{25ACT_DR6} at smaller angular scales, as well as the South Pole Telescope~(SPT)-3G~D1~\cite{25Camphuis_SPT3Gd1}, using consistent foreground modeling. Where indicated, we additionally include CMB lensing~(L) from \Planck+ACT+SPT-3G~\cite{25Ge_SPT3GMUSE} and SN data from the DES~Dovekie compilation~\cite{25Dovekie_DESY5SNe}. We sample the six standard \LCDM parameters plus \Og, assuming one massive neutrino with $m_\nu=0.06\,\mathrm{eV}$ and $N_\mathrm{eff}=3.044$. Perturbations are evolved with the parameterized post-Friedmann (PPF) dark-energy prescription in a modified \texttt{CAMB} code~\cite{00Lewis_CAMB,12Howlett_CAMB,24WenCGG}, with sound-speed robustness tests described below. Likelihoods, priors, chains, and nested-sampling evidence calculations are detailed in the Supplemental Material (\cref{app:supplement}).

\begin{figure}[t]
    \centering
    \includegraphics[width=\columnwidth]{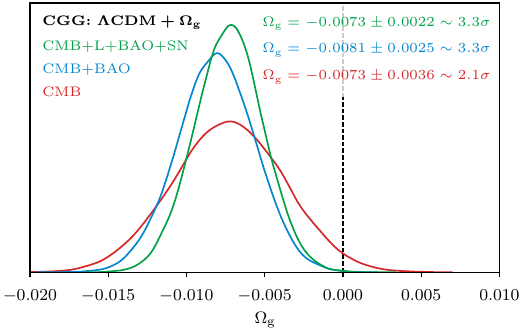}
    \caption{Marginalized posterior of the cosmic glitch parameter~\Og for primary CMB data alone (\Planck+ACT+SPT), for CMB+BAO (adding DESI~DR2), and for CMB+L+BAO+SN (further adding CMB lensing and DES~Dovekie supernovae). The dashed line marks the \LCDM value $\Og=0$.
    }
    \label{fig:Omega_g}
\end{figure}

\Cref{fig:Omega_g} shows the resulting posteriors. From CMB+BAO we obtain
\begin{equation}
    \Og = \num{-0.0081(25)},
    \label{eq:main_result}
\end{equation}
along with ${\Om=\num{0.300(4)}}$, ${r_\mathrm{d}h=\num{101.5(5)}}$, ${H_0=\SI{68.31(26)}{\km\per\s\per\mega\parsec}}$, ${\sigma_8=\num{0.836(9)}}$, and ${S_8=\num{0.835(10)}}$. The data thus prefer a negative glitch parameter at $3.3\,\sigma$. In the physical form of \cref{eq:Og}, this is
\begin{equation}
    \frac{\Gc}{\GN} = \frac{1}{1-\Og} = \num{0.9920(25)},
    \label{eq:g_ratio}
\end{equation}
a \SI{0.8}{\percent} weakening of gravity on cosmological scales, which maps through \cref{eq:horava_mapping} to
\begin{equation}
    \lambda-1 = -\tfrac{2}{3}\,\Og = \num{0.0054(17)}.
    \label{eq:lambda_result}
\end{equation}
Adding CMB lensing and supernovae, i.e., for CMB+L+BAO+SN, we find $\Og=\num{-0.0073(22)}$, corresponding to $\Gc/\GN=\num{0.9928(22)}$ and $\lambda-1=\num{0.0049(15)}$, still $3.3\,\sigma$ below zero. The primary CMB alone already prefers $\Og=\num{-0.0073(36)}$ ($2.1\,\sigma$); the BAO data sharpen the constraint without shifting its center. The progression documented in the \cref{app:supplement} shows this is not a fluctuation of one data set: a tendency toward $\Og<0$ was already present in \Planck-PR3 alone, grew with DESI~DR1, survives the DESI~DR2 update, and sharpens further when the high-resolution ACT and SPT spectra are added.

\begin{figure*}[t]
    \centering
    \includegraphics[width=\textwidth]{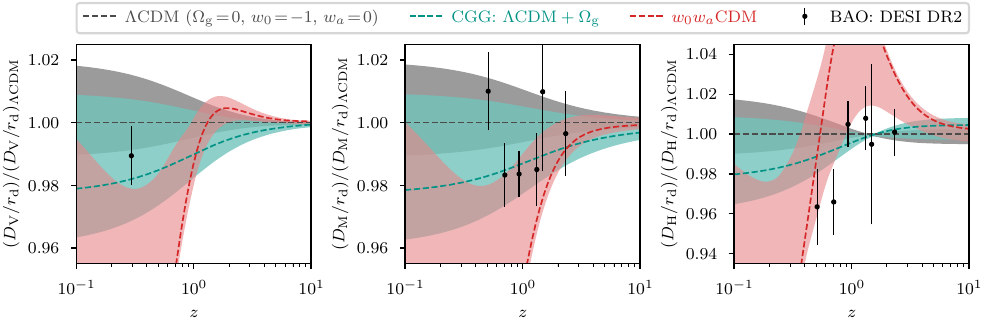}
    \caption{
        Posterior predictions for the DESI~DR2 BAO distances~\cite{25DESI_DR2BAO} from the primary CMB alone, for \LCDM (gray), CGG (teal), and \wwCDM (red). Dashed lines show the best fit to the \Hillik CMB likelihood~\cite{26Tristram_Hillik} for each model, and shaded bands the $\pm2\,\sigma$ range of the CMB posterior; all curves are ratios to the \LCDM CMB best fit. No BAO information enters the fits. The panels show $D_\mathrm{V}/r_\mathrm{d}$ (angle-averaged), $D_\mathrm{M}/r_\mathrm{d}$ (transverse), and $D_\mathrm{H}/r_\mathrm{d}$ (radial) distances; black points are the DESI tracers in increasing redshift order: BGS, LRG1, LRG2, LRG3+ELG1, ELG2, QSO, and Ly\,$\alpha$.
    }
    \label{fig:bao_residuals}
\end{figure*}

Why do the BAO data pull on \Og? \Cref{fig:bao_residuals} answers this by treating the CMB fit of each model as a \emph{prediction} for the BAO distances, rather than fitting CMB and BAO jointly. A joint fit would flatter \wwCDM, whose two extra parameters can absorb almost any smooth distance--redshift relation: at the joint CMB+BAO best fits, ${\Delta\chi^2_\mathrm{BAO}=-2.2}$ for CGG and $-5.3$ for \wwCDM relative to \LCDM (\cref{tab:bestfit}). The more demanding test is whether a model fitted to the CMB anticipates what DESI then measures. \LCDM fitted to the CMB predicts distances that sit systematically above the DESI $D_{V}/r_\mathrm{d}$ and $D_{M}/r_\mathrm{d}$ points at $z\lesssim1$~\cite{25DESI_DR2BAO}. CGG fitted to the CMB predicts a \SIrange[range-units=single,range-phrase=--]{1}{2}{\percent} downward shift of exactly this form, with a band comparable in width to the \LCDM one, and passes through the DESI points. \wwCDM fitted to the CMB alone makes essentially no prediction: its band spans several percent in either direction, because $w_0$ and $w_a$ are unconstrained without low-redshift data.
The 1-parameter glitch is therefore not merely a fit to DESI; it is a CMB-calibrated prediction that DESI confirms.

\begin{figure*}[t]
    \centering
    \includegraphics[width=0.49\textwidth]{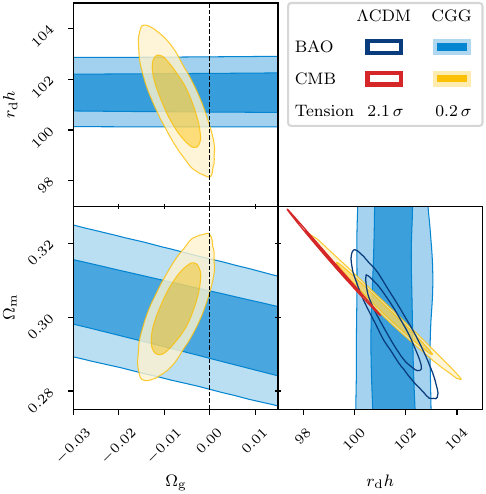}\hfill
    \includegraphics[width=0.49\textwidth]{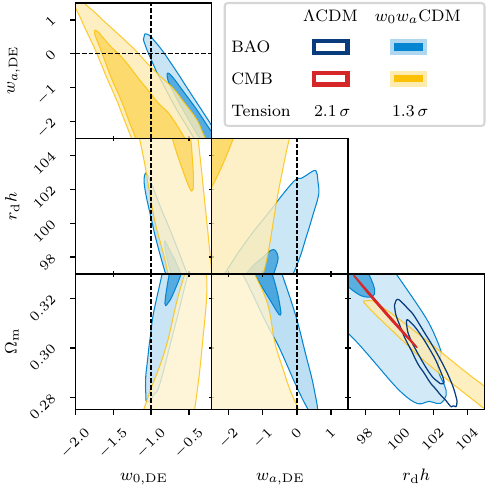}
    \caption{
        The CMB--BAO tension and its resolution.
        Separate CMB-only and BAO-only posteriors are shown in the plane of $r_\mathrm{d}h$ and \Om, for \LCDM (open contours) and for the extended model (filled contours), with their tension quoted in each legend.
        \emph{Left:} the CGG extension. Adding \Og barely widens the CMB posterior but frees the BAO posterior, and the two constraints move toward one another.
        \emph{Right:} the \wwCDM extension. Both CMB and BAO posteriors become individually nearly unconstrained, and their degeneracy directions pull apart rather than together.
    }
    \label{fig:cmb_bao_tension}
\end{figure*}

The same point can be made in parameter space. In \LCDM, the CMB and DESI~DR2 BAO data disagree at $2.1\,\sigma$ in the $(r_\mathrm{d}h,\Om)$ plane (\cref{fig:cmb_bao_tension}, open contours), the well-known preference of DESI for a lower \Om than the CMB~\cite{25DESI_DR2BAO,24Choudhury_cmb_bao,25Ye_cmb_bao,25Ong_cmb_bao,25Gu_cmb_bao,26Ferreira_cmb_bao,26Lee_cmb_bao} (see Ref.~\cite{26Hergt_BMCtension} for the tension metric). Both extensions relax this tension, but in different ways.
Under CGG the CMB constraint on $(r_\mathrm{d}h,\Om)$ remains tight, but uses the opened-up \Og space to stretch to the BAO preference for ${r_\mathrm{d}h\gtrsim100}$. The BAO constraint on $r_\mathrm{d}h$ is unchanged relative to \LCDM, but \Om becomes degenerate with \Og. As a result, BAO data alone cannot constrain either density, yet it can fully overlap with the CMB constraint, reducing the remaining tension to $0.2\,\sigma$.
Under \wwCDM, both CMB and BAO posteriors on $(r_\mathrm{d}h,\Om)$ become individually quite unconstrained, with the CMB stretching to smaller \Om and larger $r_\mathrm{d}h$ and the BAO expanding in the opposite direction, curiously inverting their \LCDM preference. The tension is reduced only to $1.3\,\sigma$. A single parameter that makes two data sets agree is more informative than two parameters that make each of them uninformative.

\begin{table}[t]
    \setlength{\tabcolsep}{3pt} %
    \small
    \centering
    \caption{\label{tab:bestfit}
        Model comparison for CMB+BAO. The columns give the number of extra parameters, the minimum $\chi^2$ improvement, and the log-evidence difference relative to \LCDM, and the marginalized constraints on the extension parameters. These runs use the foreground-marginalized \SPA \lite CMB likelihood (with the same \Planck/ACT multipole thresholds as \Hillik) and DESI~DR2, because the roughly 70 nuisance parameters of \Hillik make maximization and nested sampling prohibitively expensive; hence the CGG constraint differs slightly from \cref{eq:main_result}. Evidence uncertainties are the nested-sampling errors.
    }
    \begin{tabularx}{1.0\columnwidth}{l@{\hspace{2pt}}c c c@{}c l}
        \toprule
            Model    & No. & $\Delta\chi^2_\mathrm{CMB}$ & $\Delta\chi^2_\mathrm{BAO}$ & $\Delta\ln\mathcal{Z}$ & \multicolumn{1}{c}{Constraints} \\
        \midrule
            \LCDM    & 0  & $0$    & $0$    & $0$         &                              \\ \addlinespace[\smallskipamount]
            CGG      & 1  & $-6.4$ & $-2.2$ & $3.2\pm0.2$ & $\Og{=}{-}0.0091{\pm}0.0025$ \\ \addlinespace[\smallskipamount]
            $w_0w_a$ & 2  & $-5.6$ & $-5.3$ & $3.2\pm0.2$ & $w_0\!=\!-0.40\pm0.20$       \\
                     &    &        &        &             & $w_a\!=\!-1.8\pm0.6$         \\
        \bottomrule
    \end{tabularx}
\end{table}

The strongest reason to take CGG seriously is not only the nominal significance of \cref{eq:main_result}, but its economy (\cref{tab:bestfit}). The minimum $\chi^2$ improves from \LCDM to CGG by ${\Delta\chi^2_\mathrm{tot}=-6.4-2.2=-8.6}$, compared with $-10.9$ for \wwCDM with twice as many extra parameters. Bayesian evidence applies the corresponding Occam penalty. For the priors listed in the \cref{app:supplement}, nested sampling on CMB+BAO gives ${\Delta\ln\mathcal{Z}=+3.2}$ in favor of both CGG and \wwCDM relative to \LCDM: the 2-parameter model buys nothing over the 1-parameter glitch. This is not decisive evidence against \wwCDM and should not be read as a model-selection verdict; evidence ratios are prior dependent, and an unconstrained extension parameter incurs almost no Occam penalty, so the nominal equality in $\Delta\ln\mathcal{Z}$ understates the predictive advantage of CGG visible in \cref{fig:bao_residuals}. When CMB lensing and SN data are added, the preference for CGG over \LCDM weakens to ${\Delta\ln\mathcal{Z}=+1.1\pm0.2}$. The log-evidence for \wwCDM weakens only to ${+2.2\pm0.2}$, because its extra freedom can accommodate the higher \Om preferred by supernovae relative to both CMB and BAO~\cite{25DESI_DR2BAO}; the evidence decompositions are given in the \cref{app:supplement}.

\begin{figure}[t]
    \centering
    \includegraphics[width=\columnwidth]{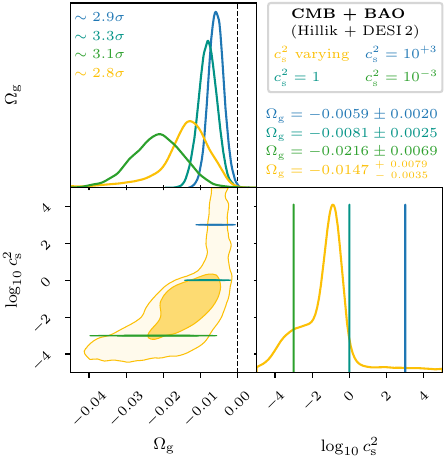}
    \caption{
        Joint posterior of the glitch parameter $\Og$ and the sound speed $\cs$ for CMB+BAO.
        For $\cs>1$ there is no correlation with \Og, but for $\cs<1$ they are positively correlated, yielding strongly skewed marginal 1D distributions.
        To assess the significance of $\Og<0$ after marginalization, we use the iso-probability (``waterlevel'') credibility interval; the standard deviation is unsuitable because the skew inflates it.
    }
    \label{fig:cs2_2d}
\end{figure}

The main theoretical caveat is perturbations. For $\Og<0$ the effective dark-energy density of \cref{eq:rho_cgg} is negative at early times and crosses zero before approaching the cosmological constant regime, so its equation-of-state description formally diverges. In CGG this is a bookkeeping singularity of the effective fluid representation, not a phantom instability of a physical scalar field.
Our fiducial PPF runs use the rest-frame sound speed squared~${\cs^2=1}$. As shown in \cref{fig:cs2_2d}, repeating the analysis with a smooth cuscuton-like proxy ${\cs^2=10^3}$ changes the central value by less than $1\,\sigma$. Marginalizing $\log_{10}\cs^2$ over $[-5,5]$ shifts the central value slightly more, mostly driven by the tail toward more negative \Og. The preference for $\Og<0$ remains in all cases at or above $2.8\,\sigma$ (\cref{fig:cs2_2d}). Although the marginal posterior in \cref{fig:cs2_2d} puts more weight at $\cs^2<1$, largely because \Og is less constrained there, this region is excluded for the khronon by the Solar-System bound of \cref{eq:cs_bound} and by the absence of gravitational Cherenkov radiation, which requires $\cs\geq1$~\cite{25KovachikSibiryakov_khronoBH}; in the allowed regime $\cs^2\gg1$ the constraint is that of the ${\cs^2=10^3}$ run, ${\Og=-0.0059\pm0.0020}$, and is independent of the precise value of \cs.
The exact $\cs\to\infty$ cuscuton/minimal-Ho\v{r}ava limit is known to coincide with this smooth-fluid description~\cite{07CuscutonCosmo,09CuscutonHovrava}; the finite-$\cs$ mapping from a microscopic Ho\v{r}ava action is left to future work.

How can the glitch be tested further? Beyond the distance sector, it leaves a distinctive imprint on the growth of structure, because the Friedmann equation is governed by \Gc while the Poisson equation retains \GN. During matter domination linear perturbations grow as ${\delta\propto a^{1-3\Og/5}}$ rather than ${\delta\propto a}$ (\cref{app:supplement}), so the gravitational potential is not frozen even before dark energy dominates: $\Phi\propto a^{-3\Og/5}$ grows slowly for $\Og<0$, producing an early-time integrated Sachs--Wolfe contribution absent in \LCDM. The main effect is cumulative: linear growth since recombination is enhanced by ${(1+z_\ast)^{-3\Og/5}\simeq1.035}$, a lever that grows rapidly with $|\Og|$. In the CMB fit, part of this is absorbed along the $A_\mathrm{s}$--$\tau$ degeneracy, which keeps $A_\mathrm{s}e^{-2\tau}$ fixed, leaving $\sigma_8$ higher by \SI{2.8}{\percent} (\cref{app:supplement}). A direct test is therefore the ratio of the low-redshift clustering amplitude to that inferred from the primary CMB, which CGG predicts to exceed the \LCDM extrapolation by a few percent. Because the preferred \Om shifts along with \Og~\cite{24WenCGG}, this comparison requires analyzing the low-redshift data within CGG itself; percent-level amplitudes from Euclid and Rubin will make it sharp. In standard modified-gravity phenomenology this is a scale-independent shift of the growth parameter $\mu$ with no gravitational slip, $\Phi=\Psi$, so lensing and dynamics respond in lockstep, a pattern that distinguishes CGG from most scalar--tensor alternatives. At late times this appears as a nearly redshift-independent roughly \SI{3}{\percent} increase in $f\sigma_8$ over ${0.3<z<1.5}$, carried by the amplitude rather than by $f$ itself, a clean target for DESI and Euclid redshift-space-distortion and full-shape analyses, and the evolving potential is testable through ISW--galaxy cross-correlations and CMB lensing from ACT, SPT, and the Simons Observatory. Since nonlinear dynamics remain Newtonian and only the expansion history changes, $N$-body simulations of CGG should be straightforward and could extend these predictions to the nonlinear regime, while a first-principles treatment of khronon perturbations at finite \cs will sharpen the theoretical target.

We therefore arrive at a compact interpretation of the post-DESI data. If the data are fit with evolving dark energy, they ask for two parameters whose physical origin is unspecified and which, on their own, predict nothing about the BAO scale. If they are fit with a cosmological-to-Newtonian gravitational mismatch, they ask for one parameter, $\Gc/\GN=0.9920\pm0.0025$, calibrated on the CMB and confirmed by DESI. This is evidence for a percent-level departure from the assumption that the cosmological and Newtonian gravitational couplings are identical, in the direction realized by Ho\v{r}ava-like and cuscuton-like modified gravity: in the normalization of \cref{eq:horava_mapping}, the data prefer a positive Ho\v{r}ava kinetic deformation at the half-percent level. It is not a detection of Ho\v{r}ava gravity as a complete theory, but it makes definite predictions for growth, lensing, and the ISW effect that forthcoming surveys can test. The result thus changes the phenomenological target: the DESI-era anomaly need not be read first as dynamical dark energy. By Occam's razor, a 1-parameter glitch in gravity deserves to be tested alongside, and not after, $w_0w_a$.

\section*{Acknowledgements}

NA thanks Sergey Sibiryakov and Shinji Mukohayama for helpful discussions. 

This research was supported by the Natural Sciences and Engineering Research Council of Canada.
LTH is supported by a Postdoctoral Fellowship from the Centre National de la Recherche Scientifique~(CNRS) in France.
NA is further supported by the Perimeter Institute for Theoretical Physics; research at the Perimeter Institute is supported in part by the Government of Canada through the Department of Innovation, Science and Economic Development Canada and by the Province of Ontario through the Ministry of Colleges and Universities.
We further gratefully acknowledge support from the CNRS/IN2P3 Computing Center (Lyon -- France) for providing computing and data-processing resources needed for this work.

Parts of this work use CMB observations from Planck, ACT, and SPT-3G; BAO measurements from DESI; and SN data from the Dark Energy Survey, the Pantheon+ team, and the Supernova Cosmology Project. We thank the teams and collaborations for making these data publicly available.
This paper made use of codes including \href{https://camb.readthedocs.io/}{\texttt{CAMB}}, \href{https://cobaya.readthedocs.io/}{\texttt{Cobaya}}, \href{https://github.com/PolyChord/PolyChordLite}{\texttt{PolyChord}}, and \href{https://anesthetic.readthedocs.io/}{\texttt{anesthetic}}.

\section*{Data availability}
The data that support the findings of this Letter (MCMC and nested sampling chains, post-processing and visualization code) are openly available on \textsc{Zenodo}~\cite{Zenodo2026_hergt_glitch}. The modified \texttt{CAMB} version can be found at \url{https://github.com/lukashergt/CAMB/tree/CGG}.

\bibliographystyle{apsrev4-2}
\bibliography{refs}

\begin{thebibliography}{57}%
\makeatletter
\providecommand \@ifxundefined [1]{%
 \@ifx{#1\undefined}
}%
\providecommand \@ifnum [1]{%
 \ifnum #1\expandafter \@firstoftwo
 \else \expandafter \@secondoftwo
 \fi
}%
\providecommand \@ifx [1]{%
 \ifx #1\expandafter \@firstoftwo
 \else \expandafter \@secondoftwo
 \fi
}%
\providecommand \natexlab [1]{#1}%
\providecommand \enquote  [1]{``#1''}%
\providecommand \bibnamefont  [1]{#1}%
\providecommand \bibfnamefont [1]{#1}%
\providecommand \citenamefont [1]{#1}%
\providecommand \href@noop [0]{\@secondoftwo}%
\providecommand \href [0]{\begingroup \@sanitize@url \@href}%
\providecommand \@href[1]{\@@startlink{#1}\@@href}%
\providecommand \@@href[1]{\endgroup#1\@@endlink}%
\providecommand \@sanitize@url [0]{\catcode `\\12\catcode `\$12\catcode `\&12\catcode `\#12\catcode `\^12\catcode `\_12\catcode `\%12\relax}%
\providecommand \@@startlink[1]{}%
\providecommand \@@endlink[0]{}%
\providecommand \url  [0]{\begingroup\@sanitize@url \@url }%
\providecommand \@url [1]{\endgroup\@href {#1}{\urlprefix }}%
\providecommand \urlprefix  [0]{URL }%
\providecommand \Eprint [0]{\href }%
\providecommand \doibase [0]{https://doi.org/}%
\providecommand \selectlanguage [0]{\@gobble}%
\providecommand \bibinfo  [0]{\@secondoftwo}%
\providecommand \bibfield  [0]{\@secondoftwo}%
\providecommand \translation [1]{[#1]}%
\providecommand \BibitemOpen [0]{}%
\providecommand \bibitemStop [0]{}%
\providecommand \bibitemNoStop [0]{.\EOS\space}%
\providecommand \EOS [0]{\spacefactor3000\relax}%
\providecommand \BibitemShut  [1]{\csname bibitem#1\endcsname}%
\let\auto@bib@innerbib\@empty
\bibitem [{\citenamefont {{DESI Collaboration}}(2025)}]{25DESI_DR2BAO}%
  \BibitemOpen
  \bibfield  {author} {\bibinfo {author} {\bibnamefont {{DESI Collaboration}}},\ }\href {https://doi.org/10.1103/tr6y-kpc6} {\bibfield  {journal} {\bibinfo  {journal} {\prd}\ }\textbf {\bibinfo {volume} {112}},\ \bibinfo {eid} {083515} (\bibinfo {year} {2025})},\ \Eprint {https://arxiv.org/abs/2503.14738} {arXiv:2503.14738 [astro-ph.CO]} \BibitemShut {NoStop}%
\bibitem [{\citenamefont {{SPT-3G Collaboration}}\ \emph {et~al.}(2026)\citenamefont {{SPT-3G Collaboration}}, \citenamefont {{Camphuis}}, \citenamefont {{Quan}}, \citenamefont {{Balkenhol}}, \citenamefont {{Khalife}}, \citenamefont {{Ge}}, \citenamefont {{Guidi}}, \citenamefont {{Huang}}, \citenamefont {{Lynch}}, \citenamefont {{Omori}}, \citenamefont {{Trendafilova}},\ and\ \citenamefont {{et al.}}}]{25Camphuis_SPT3Gd1}%
  \BibitemOpen
  \bibfield  {author} {\bibinfo {author} {\bibnamefont {{SPT-3G Collaboration}}}, \bibinfo {author} {\bibfnamefont {E.}~\bibnamefont {{Camphuis}}}, \bibinfo {author} {\bibfnamefont {W.}~\bibnamefont {{Quan}}}, \emph {et~al.},\ }\href {https://doi.org/10.1103/7wt3-9v2y} {\bibfield  {journal} {\bibinfo  {journal} {\prd}\ }\textbf {\bibinfo {volume} {113}},\ \bibinfo {eid} {083504} (\bibinfo {year} {2026})},\ \Eprint {https://arxiv.org/abs/2506.20707} {arXiv:2506.20707 [astro-ph.CO]} \BibitemShut {NoStop}%
\bibitem [{\citenamefont {{Herold}}\ and\ \citenamefont {{Karwal}}(2026)}]{25HeroldKarwal_DDE}%
  \BibitemOpen
  \bibfield  {author} {\bibinfo {author} {\bibfnamefont {L.}~\bibnamefont {{Herold}}}\ and\ \bibinfo {author} {\bibfnamefont {T.}~\bibnamefont {{Karwal}}},\ }\href {https://doi.org/10.1103/gw6q-5k5j} {\bibfield  {journal} {\bibinfo  {journal} {\prd}\ }\textbf {\bibinfo {volume} {113}},\ \bibinfo {eid} {123551} (\bibinfo {year} {2026})},\ \Eprint {https://arxiv.org/abs/2506.12004} {arXiv:2506.12004 [astro-ph.CO]} \BibitemShut {NoStop}%
\bibitem [{\citenamefont {{Li}}\ \emph {et~al.}(2026)\citenamefont {{Li}}, \citenamefont {{Du}}, \citenamefont {{Zhou}}, \citenamefont {{Li}}, \citenamefont {{Zhang}},\ and\ \citenamefont {{Zhang}}}]{26Li_DDE_ACTSPTPlanck}%
  \BibitemOpen
  \bibfield  {author} {\bibinfo {author} {\bibfnamefont {T.-N.}\ \bibnamefont {{Li}}}, \bibinfo {author} {\bibfnamefont {G.-H.}\ \bibnamefont {{Du}}}, \bibinfo {author} {\bibfnamefont {S.-H.}\ \bibnamefont {{Zhou}}}, \emph {et~al.},\ }\href {https://doi.org/10.1016/j.dark.2026.102254} {\bibfield  {journal} {\bibinfo  {journal} {Physics of the Dark Universe}\ }\textbf {\bibinfo {volume} {52}},\ \bibinfo {eid} {102254} (\bibinfo {year} {2026})},\ \Eprint {https://arxiv.org/abs/2511.22512} {arXiv:2511.22512 [astro-ph.CO]} \BibitemShut {NoStop}%
\bibitem [{\citenamefont {Hergt}\ \emph {et~al.}(2026)\citenamefont {Hergt}, \citenamefont {Henrot-Versill{\'e}}, \citenamefont {Tristram},\ and\ \citenamefont {Scott}}]{26Hergt_BMCtension}%
  \BibitemOpen
  \bibfield  {author} {\bibinfo {author} {\bibfnamefont {L.~T.}\ \bibnamefont {Hergt}}, \bibinfo {author} {\bibfnamefont {S.}~\bibnamefont {Henrot-Versill{\'e}}}, \bibinfo {author} {\bibfnamefont {M.}~\bibnamefont {Tristram}},\ and\ \bibinfo {author} {\bibfnamefont {D.}~\bibnamefont {Scott}},\ }\href@noop {} {\bibfield  {journal} {\bibinfo  {journal} {Phys. Rev. D}\ } (\bibinfo {year} {2026})},\ \bibinfo {note} {in press},\ \Eprint {https://arxiv.org/abs/2602.06115} {arXiv:2602.06115} \BibitemShut {NoStop}%
\bibitem [{\citenamefont {Chevallier}\ and\ \citenamefont {Polarski}(2001)}]{01Chevallier_CPL}%
  \BibitemOpen
  \bibfield  {author} {\bibinfo {author} {\bibfnamefont {M.}~\bibnamefont {Chevallier}}\ and\ \bibinfo {author} {\bibfnamefont {D.}~\bibnamefont {Polarski}},\ }\href {https://doi.org/10.1142/S0218271801000822} {\bibfield  {journal} {\bibinfo  {journal} {Int. J. Mod. Phys. D}\ }\textbf {\bibinfo {volume} {10}},\ \bibinfo {pages} {213} (\bibinfo {year} {2001})},\ \Eprint {https://arxiv.org/abs/gr-qc/0009008} {arXiv:gr-qc/0009008} \BibitemShut {NoStop}%
\bibitem [{\citenamefont {Linder}(2003)}]{03Linder_CPL}%
  \BibitemOpen
  \bibfield  {author} {\bibinfo {author} {\bibfnamefont {E.~V.}\ \bibnamefont {Linder}},\ }\href {https://doi.org/10.1103/PhysRevLett.90.091301} {\bibfield  {journal} {\bibinfo  {journal} {Phys. Rev. Lett.}\ }\textbf {\bibinfo {volume} {90}},\ \bibinfo {pages} {091301} (\bibinfo {year} {2003})},\ \Eprint {https://arxiv.org/abs/astro-ph/0208512} {arXiv:astro-ph/0208512} \BibitemShut {NoStop}%
\bibitem [{\citenamefont {{Wen}}\ \emph {et~al.}(2024)\citenamefont {{Wen}}, \citenamefont {{Hergt}}, \citenamefont {{Afshordi}},\ and\ \citenamefont {{Scott}}}]{24WenCGG}%
  \BibitemOpen
  \bibfield  {author} {\bibinfo {author} {\bibfnamefont {R.~Y.}\ \bibnamefont {{Wen}}}, \bibinfo {author} {\bibfnamefont {L.~T.}\ \bibnamefont {{Hergt}}}, \bibinfo {author} {\bibfnamefont {N.}~\bibnamefont {{Afshordi}}},\ and\ \bibinfo {author} {\bibfnamefont {D.}~\bibnamefont {{Scott}}},\ }\href {https://doi.org/10.1088/1475-7516/2024/03/045} {\bibfield  {journal} {\bibinfo  {journal} {\jcap}\ }\textbf {\bibinfo {volume} {2024}},\ \bibinfo {eid} {045} (\bibinfo {year} {2024})},\ \Eprint {https://arxiv.org/abs/2311.03028} {arXiv:2311.03028 [astro-ph.CO]} \BibitemShut {NoStop}%
\bibitem [{\citenamefont {Wen}\ \emph {et~al.}(2025)\citenamefont {Wen}, \citenamefont {Hergt}, \citenamefont {Afshordi},\ and\ \citenamefont {Scott}}]{24WenCGG_essay}%
  \BibitemOpen
  \bibfield  {author} {\bibinfo {author} {\bibfnamefont {R.~Y.}\ \bibnamefont {Wen}}, \bibinfo {author} {\bibfnamefont {L.~T.}\ \bibnamefont {Hergt}}, \bibinfo {author} {\bibfnamefont {N.}~\bibnamefont {Afshordi}},\ and\ \bibinfo {author} {\bibfnamefont {D.}~\bibnamefont {Scott}},\ }\href {https://doi.org/10.1007/978-3-031-90186-7_35} {\bibfield  {journal} {\bibinfo  {journal} {Astrophys. Space Sci. Proc.}\ }\textbf {\bibinfo {volume} {61}},\ \bibinfo {pages} {435} (\bibinfo {year} {2025})},\ \Eprint {https://arxiv.org/abs/2412.09568} {arXiv:2412.09568 [astro-ph.CO]} \BibitemShut {NoStop}%
\bibitem [{\citenamefont {Robbers}\ \emph {et~al.}(2008)\citenamefont {Robbers}, \citenamefont {Afshordi},\ and\ \citenamefont {Doran}}]{Robbers:2007ca}%
  \BibitemOpen
  \bibfield  {author} {\bibinfo {author} {\bibfnamefont {G.}~\bibnamefont {Robbers}}, \bibinfo {author} {\bibfnamefont {N.}~\bibnamefont {Afshordi}},\ and\ \bibinfo {author} {\bibfnamefont {M.}~\bibnamefont {Doran}},\ }\href {https://doi.org/10.1103/PhysRevLett.100.111101} {\bibfield  {journal} {\bibinfo  {journal} {Phys. Rev. Lett.}\ }\textbf {\bibinfo {volume} {100}},\ \bibinfo {pages} {111101} (\bibinfo {year} {2008})},\ \Eprint {https://arxiv.org/abs/0708.3235} {arXiv:0708.3235 [astro-ph]} \BibitemShut {NoStop}%
\bibitem [{\citenamefont {{Ho{\v{r}}ava}}(2009)}]{09Hovrava}%
  \BibitemOpen
  \bibfield  {author} {\bibinfo {author} {\bibfnamefont {P.}~\bibnamefont {{Ho{\v{r}}ava}}},\ }\href {https://doi.org/10.1103/PhysRevD.79.084008} {\bibfield  {journal} {\bibinfo  {journal} {\prd}\ }\textbf {\bibinfo {volume} {79}},\ \bibinfo {eid} {084008} (\bibinfo {year} {2009})},\ \Eprint {https://arxiv.org/abs/0901.3775} {arXiv:0901.3775 [hep-th]} \BibitemShut {NoStop}%
\bibitem [{\citenamefont {{Jacobson}}(2010)}]{10Jacobson_Aether}%
  \BibitemOpen
  \bibfield  {author} {\bibinfo {author} {\bibfnamefont {T.}~\bibnamefont {{Jacobson}}},\ }\href {https://doi.org/10.1103/PhysRevD.81.101502} {\bibfield  {journal} {\bibinfo  {journal} {\prd}\ }\textbf {\bibinfo {volume} {81}},\ \bibinfo {eid} {101502} (\bibinfo {year} {2010})},\ \Eprint {https://arxiv.org/abs/1001.4823} {arXiv:1001.4823 [hep-th]} \BibitemShut {NoStop}%
\bibitem [{\citenamefont {{Afshordi}}\ \emph {et~al.}(2007)\citenamefont {{Afshordi}}, \citenamefont {{Chung}}, \citenamefont {{Doran}},\ and\ \citenamefont {{Geshnizjani}}}]{07CuscutonCosmo}%
  \BibitemOpen
  \bibfield  {author} {\bibinfo {author} {\bibfnamefont {N.}~\bibnamefont {{Afshordi}}}, \bibinfo {author} {\bibfnamefont {D.~J.~H.}\ \bibnamefont {{Chung}}}, \bibinfo {author} {\bibfnamefont {M.}~\bibnamefont {{Doran}}},\ and\ \bibinfo {author} {\bibfnamefont {G.}~\bibnamefont {{Geshnizjani}}},\ }\href {https://doi.org/10.1103/PhysRevD.75.123509} {\bibfield  {journal} {\bibinfo  {journal} {\prd}\ }\textbf {\bibinfo {volume} {75}},\ \bibinfo {eid} {123509} (\bibinfo {year} {2007})},\ \Eprint {https://arxiv.org/abs/astro-ph/0702002} {arXiv:astro-ph/0702002 [astro-ph]} \BibitemShut {NoStop}%
\bibitem [{\citenamefont {{Afshordi}}(2009)}]{09CuscutonHovrava}%
  \BibitemOpen
  \bibfield  {author} {\bibinfo {author} {\bibfnamefont {N.}~\bibnamefont {{Afshordi}}},\ }\href {https://doi.org/10.1103/PhysRevD.80.081502} {\bibfield  {journal} {\bibinfo  {journal} {\prd}\ }\textbf {\bibinfo {volume} {80}},\ \bibinfo {eid} {081502} (\bibinfo {year} {2009})},\ \Eprint {https://arxiv.org/abs/0907.5201} {arXiv:0907.5201 [hep-th]} \BibitemShut {NoStop}%
\bibitem [{\citenamefont {{Pons}}\ and\ \citenamefont {{Talavera}}(2010)}]{10Pons_minimal_Horava}%
  \BibitemOpen
  \bibfield  {author} {\bibinfo {author} {\bibfnamefont {J.~M.}\ \bibnamefont {{Pons}}}\ and\ \bibinfo {author} {\bibfnamefont {P.}~\bibnamefont {{Talavera}}},\ }\href {https://doi.org/10.1103/PhysRevD.82.044011} {\bibfield  {journal} {\bibinfo  {journal} {\prd}\ }\textbf {\bibinfo {volume} {82}},\ \bibinfo {eid} {044011} (\bibinfo {year} {2010})},\ \Eprint {https://arxiv.org/abs/1003.3811} {arXiv:1003.3811 [gr-qc]} \BibitemShut {NoStop}%
\bibitem [{\citenamefont {{Jacobson}}\ and\ \citenamefont {{Pulakkat}}(2025)}]{25JacobsonPulakkat_mHg}%
  \BibitemOpen
  \bibfield  {author} {\bibinfo {author} {\bibfnamefont {T.}~\bibnamefont {{Jacobson}}}\ and\ \bibinfo {author} {\bibfnamefont {P.}~\bibnamefont {{Pulakkat}}},\ }\href {https://doi.org/10.1088/1751-8121/ade918} {\bibfield  {journal} {\bibinfo  {journal} {J. Phys. A: Math. Theor.}\ }\textbf {\bibinfo {volume} {58}},\ \bibinfo {pages} {315404} (\bibinfo {year} {2025})},\ \Eprint {https://arxiv.org/abs/2508.03106} {arXiv:2508.03106 [gr-qc]} \BibitemShut {NoStop}%
\bibitem [{\citenamefont {{Alvey}}\ \emph {et~al.}(2020)\citenamefont {{Alvey}}, \citenamefont {{Sabti}}, \citenamefont {{Escudero}},\ and\ \citenamefont {{Fairbairn}}}]{20Alvey_BBN_G}%
  \BibitemOpen
  \bibfield  {author} {\bibinfo {author} {\bibfnamefont {J.}~\bibnamefont {{Alvey}}}, \bibinfo {author} {\bibfnamefont {N.}~\bibnamefont {{Sabti}}}, \bibinfo {author} {\bibfnamefont {M.}~\bibnamefont {{Escudero}}},\ and\ \bibinfo {author} {\bibfnamefont {M.}~\bibnamefont {{Fairbairn}}},\ }\href {https://doi.org/10.1140/epjc/s10052-020-7727-y} {\bibfield  {journal} {\bibinfo  {journal} {Eur. Phys. J. C}\ }\textbf {\bibinfo {volume} {80}},\ \bibinfo {eid} {148} (\bibinfo {year} {2020})},\ \Eprint {https://arxiv.org/abs/1910.10730} {arXiv:1910.10730 [astro-ph.CO]} \BibitemShut {NoStop}%
\bibitem [{\citenamefont {{Yagi}}\ \emph {et~al.}(2014)\citenamefont {{Yagi}}, \citenamefont {{Blas}}, \citenamefont {{Barausse}},\ and\ \citenamefont {{Yunes}}}]{14Yagi_HoravaPulsars}%
  \BibitemOpen
  \bibfield  {author} {\bibinfo {author} {\bibfnamefont {K.}~\bibnamefont {{Yagi}}}, \bibinfo {author} {\bibfnamefont {D.}~\bibnamefont {{Blas}}}, \bibinfo {author} {\bibfnamefont {E.}~\bibnamefont {{Barausse}}},\ and\ \bibinfo {author} {\bibfnamefont {N.}~\bibnamefont {{Yunes}}},\ }\href {https://doi.org/10.1103/PhysRevD.89.084067} {\bibfield  {journal} {\bibinfo  {journal} {\prd}\ }\textbf {\bibinfo {volume} {89}},\ \bibinfo {eid} {084067} (\bibinfo {year} {2014})},\ \Eprint {https://arxiv.org/abs/1311.7144} {arXiv:1311.7144 [gr-qc]} \BibitemShut {NoStop}%
\bibitem [{\citenamefont {{Frusciante}}\ and\ \citenamefont {{Benetti}}(2021)}]{21Frusciante_HoravaGW}%
  \BibitemOpen
  \bibfield  {author} {\bibinfo {author} {\bibfnamefont {N.}~\bibnamefont {{Frusciante}}}\ and\ \bibinfo {author} {\bibfnamefont {M.}~\bibnamefont {{Benetti}}},\ }\href {https://doi.org/10.1103/PhysRevD.103.104060} {\bibfield  {journal} {\bibinfo  {journal} {\prd}\ }\textbf {\bibinfo {volume} {103}},\ \bibinfo {eid} {104060} (\bibinfo {year} {2021})},\ \Eprint {https://arxiv.org/abs/2005.14705} {arXiv:2005.14705 [astro-ph.CO]} \BibitemShut {NoStop}%
\bibitem [{\citenamefont {{Oost}}\ \emph {et~al.}(2018)\citenamefont {{Oost}}, \citenamefont {{Mukohyama}},\ and\ \citenamefont {{Wang}}}]{18Oost_AetherGW170817}%
  \BibitemOpen
  \bibfield  {author} {\bibinfo {author} {\bibfnamefont {J.}~\bibnamefont {{Oost}}}, \bibinfo {author} {\bibfnamefont {S.}~\bibnamefont {{Mukohyama}}},\ and\ \bibinfo {author} {\bibfnamefont {A.}~\bibnamefont {{Wang}}},\ }\href {https://doi.org/10.1103/PhysRevD.97.124023} {\bibfield  {journal} {\bibinfo  {journal} {\prd}\ }\textbf {\bibinfo {volume} {97}},\ \bibinfo {eid} {124023} (\bibinfo {year} {2018})},\ \Eprint {https://arxiv.org/abs/1802.04303} {arXiv:1802.04303 [gr-qc]} \BibitemShut {NoStop}%
\bibitem [{\citenamefont {{Kovachik}}\ and\ \citenamefont {{Sibiryakov}}(2025)}]{25KovachikSibiryakov_khronoBH}%
  \BibitemOpen
  \bibfield  {author} {\bibinfo {author} {\bibfnamefont {A.}~\bibnamefont {{Kovachik}}}\ and\ \bibinfo {author} {\bibfnamefont {S.}~\bibnamefont {{Sibiryakov}}},\ }\href {https://doi.org/10.1103/PhysRevD.111.044042} {\bibfield  {journal} {\bibinfo  {journal} {\prd}\ }\textbf {\bibinfo {volume} {111}},\ \bibinfo {eid} {044042} (\bibinfo {year} {2025})},\ \Eprint {https://arxiv.org/abs/2311.12936} {arXiv:2311.12936 [gr-qc]} \BibitemShut {NoStop}%
\bibitem [{\citenamefont {{Abbott}}\ \emph {et~al.}(2017)\citenamefont {{Abbott}} \emph {et~al.}}]{17Abbott_GWGRBspeed}%
  \BibitemOpen
  \bibfield  {author} {\bibinfo {author} {\bibfnamefont {B.~P.}\ \bibnamefont {{Abbott}}} \emph {et~al.},\ }\href {https://doi.org/10.3847/2041-8213/aa920c} {\bibfield  {journal} {\bibinfo  {journal} {\apjl}\ }\textbf {\bibinfo {volume} {848}},\ \bibinfo {eid} {L13} (\bibinfo {year} {2017})},\ \Eprint {https://arxiv.org/abs/1710.05834} {arXiv:1710.05834 [astro-ph.HE]} \BibitemShut {NoStop}%
\bibitem [{\citenamefont {{Gupta}}\ \emph {et~al.}(2021)\citenamefont {{Gupta}}, \citenamefont {{Herrero-Valea}}, \citenamefont {{Blas}}, \citenamefont {{Barausse}}, \citenamefont {{Cornish}}, \citenamefont {{Yagi}},\ and\ \citenamefont {{Yunes}}}]{21Gupta_AetherPulsars}%
  \BibitemOpen
  \bibfield  {author} {\bibinfo {author} {\bibfnamefont {T.}~\bibnamefont {{Gupta}}}, \bibinfo {author} {\bibfnamefont {M.}~\bibnamefont {{Herrero-Valea}}}, \bibinfo {author} {\bibfnamefont {D.}~\bibnamefont {{Blas}}}, \emph {et~al.},\ }\href {https://doi.org/10.1088/1361-6382/ac1a69} {\bibfield  {journal} {\bibinfo  {journal} {Classical and Quantum Gravity}\ }\textbf {\bibinfo {volume} {38}},\ \bibinfo {eid} {195003} (\bibinfo {year} {2021})},\ \Eprint {https://arxiv.org/abs/2104.04596} {arXiv:2104.04596 [gr-qc]} \BibitemShut {NoStop}%
\bibitem [{\citenamefont {{Ramos}}\ and\ \citenamefont {{Barausse}}(2019)}]{19Ramos_HoravaBH}%
  \BibitemOpen
  \bibfield  {author} {\bibinfo {author} {\bibfnamefont {O.}~\bibnamefont {{Ramos}}}\ and\ \bibinfo {author} {\bibfnamefont {E.}~\bibnamefont {{Barausse}}},\ }\href {https://doi.org/10.1103/PhysRevD.99.024034} {\bibfield  {journal} {\bibinfo  {journal} {\prd}\ }\textbf {\bibinfo {volume} {99}},\ \bibinfo {eid} {024034} (\bibinfo {year} {2019})},\ \Eprint {https://arxiv.org/abs/1811.07786} {arXiv:1811.07786 [gr-qc]} \BibitemShut {NoStop}%
\bibitem [{\citenamefont {{Tristram}}\ \emph {et~al.}(2026)\citenamefont {{Tristram}}, \citenamefont {{Douspis}}, \citenamefont {{Gorce}}, \citenamefont {{Henrot-Versill{\'e}}}, \citenamefont {{Hergt}}, \citenamefont {{Ilic}}, \citenamefont {{McBride}}, \citenamefont {{Mu{\~n}oz-Echeverr{\'\i}a}}, \citenamefont {{Pointecouteau}},\ and\ \citenamefont {{Salvati}}}]{26Tristram_Hillik}%
  \BibitemOpen
  \bibfield  {author} {\bibinfo {author} {\bibfnamefont {M.}~\bibnamefont {{Tristram}}}, \bibinfo {author} {\bibfnamefont {M.}~\bibnamefont {{Douspis}}}, \bibinfo {author} {\bibfnamefont {A.}~\bibnamefont {{Gorce}}}, \emph {et~al.},\ }\href {https://doi.org/10.1051/0004-6361/202558015} {\bibfield  {journal} {\bibinfo  {journal} {\aap}\ }\textbf {\bibinfo {volume} {710}},\ \bibinfo {eid} {A165} (\bibinfo {year} {2026})},\ \Eprint {https://arxiv.org/abs/2511.04733} {arXiv:2511.04733 [astro-ph.CO]} \BibitemShut {NoStop}%
\bibitem [{\citenamefont {{Planck Collaboration}}(2020{\natexlab{a}})}]{20Npipe}%
  \BibitemOpen
  \bibfield  {author} {\bibinfo {author} {\bibnamefont {{Planck Collaboration}}},\ }\href {https://doi.org/10.1051/0004-6361/202038073} {\bibfield  {journal} {\bibinfo  {journal} {\aap}\ }\textbf {\bibinfo {volume} {643}},\ \bibinfo {eid} {A42} (\bibinfo {year} {2020}{\natexlab{a}})},\ \Eprint {https://arxiv.org/abs/2007.04997} {arXiv:2007.04997 [astro-ph.CO]} \BibitemShut {NoStop}%
\bibitem [{\citenamefont {{Tristram}}\ \emph {et~al.}(2024)\citenamefont {{Tristram}}, \citenamefont {{Banday}}, \citenamefont {{Douspis}}, \citenamefont {{Garrido}}, \citenamefont {{G{\'o}rski}}, \citenamefont {{Henrot-Versill{\'e}}}, \citenamefont {{Ili{\'c}}}, \citenamefont {{Keskitalo}}, \citenamefont {{Lagache}}, \citenamefont {{Lawrence}}, \citenamefont {{Partridge}},\ and\ \citenamefont {{Scott}}}]{24Tristram_PR4}%
  \BibitemOpen
  \bibfield  {author} {\bibinfo {author} {\bibfnamefont {M.}~\bibnamefont {{Tristram}}}, \bibinfo {author} {\bibfnamefont {A.~J.}\ \bibnamefont {{Banday}}}, \bibinfo {author} {\bibfnamefont {M.}~\bibnamefont {{Douspis}}}, \emph {et~al.},\ }\href {https://doi.org/10.1051/0004-6361/202348015} {\bibfield  {journal} {\bibinfo  {journal} {\aap}\ }\textbf {\bibinfo {volume} {682}},\ \bibinfo {pages} {A37} (\bibinfo {year} {2024})},\ \Eprint {https://arxiv.org/abs/2309.10034} {arXiv:2309.10034 [astro-ph.CO]} \BibitemShut {NoStop}%
\bibitem [{\citenamefont {{The Atacama Cosmology Telescope collaboration}}\ \emph {et~al.}(2025)\citenamefont {{The Atacama Cosmology Telescope collaboration}}, \citenamefont {{Louis}}, \citenamefont {{La Posta}}, \citenamefont {{Atkins}}, \citenamefont {{Jense}},\ and\ \citenamefont {{et al.}}}]{25ACT_DR6}%
  \BibitemOpen
  \bibfield  {author} {\bibinfo {author} {\bibnamefont {{The Atacama Cosmology Telescope collaboration}}}, \bibinfo {author} {\bibfnamefont {T.}~\bibnamefont {{Louis}}}, \bibinfo {author} {\bibfnamefont {A.}~\bibnamefont {{La Posta}}}, \emph {et~al.},\ }\href {https://doi.org/10.1088/1475-7516/2025/11/062} {\bibfield  {journal} {\bibinfo  {journal} {\jcap}\ }\textbf {\bibinfo {volume} {2025}},\ \bibinfo {eid} {062} (\bibinfo {year} {2025})},\ \Eprint {https://arxiv.org/abs/2503.14452} {arXiv:2503.14452 [astro-ph.CO]} \BibitemShut {NoStop}%
\bibitem [{\citenamefont {{SPT-3G Collaboration}}\ \emph {et~al.}(2025)\citenamefont {{SPT-3G Collaboration}}, \citenamefont {{Ge}}, \citenamefont {{Millea}}, \citenamefont {{Camphuis}}, \citenamefont {{Daley}}, \citenamefont {{Huang}}, \citenamefont {{Omori}}, \citenamefont {{Quan}},\ and\ \citenamefont {{et al.}}}]{25Ge_SPT3GMUSE}%
  \BibitemOpen
  \bibfield  {author} {\bibinfo {author} {\bibnamefont {{SPT-3G Collaboration}}}, \bibinfo {author} {\bibfnamefont {F.}~\bibnamefont {{Ge}}}, \bibinfo {author} {\bibfnamefont {M.}~\bibnamefont {{Millea}}}, \emph {et~al.},\ }\href {https://doi.org/10.1103/PhysRevD.111.083534} {\bibfield  {journal} {\bibinfo  {journal} {\prd}\ }\textbf {\bibinfo {volume} {111}},\ \bibinfo {eid} {083534} (\bibinfo {year} {2025})},\ \Eprint {https://arxiv.org/abs/2411.06000} {arXiv:2411.06000 [astro-ph.CO]} \BibitemShut {NoStop}%
\bibitem [{\citenamefont {{Popovic}}\ \emph {et~al.}(2026)\citenamefont {{Popovic}}, \citenamefont {{Shah}}, \citenamefont {{Kenworthy}}, \citenamefont {{Kessler}}, \citenamefont {{Davis}}, \citenamefont {{Goobar}}, \citenamefont {{Scolnic}}, \citenamefont {{Vincenzi}}, \citenamefont {{Wiseman}}, \citenamefont {{Chen}}, \citenamefont {{Charleton}}, \citenamefont {{Acevedo}}, \citenamefont {{Armstrong}}, \citenamefont {{Boyd}}, \citenamefont {{Brout}}, \citenamefont {{Camilleri}}, \citenamefont {{Frieman}}, \citenamefont {{Galbany}}, \citenamefont {{Grayling}}, \citenamefont {{Kelsey}}, \citenamefont {{Rose}}, \citenamefont {{S{\'a}nchez}}, \citenamefont {{Lee}}, \citenamefont {{M{\"o}ller}}, \citenamefont {{Smith}}, \citenamefont {{Sullivan}}, \citenamefont {{Shiamtanis}}, \citenamefont {{Alarcon}}, \citenamefont {{Allam}}, \citenamefont {{Andrade-Oliveira}}, \citenamefont {{Avila}}, \citenamefont {{Bacon}}, \citenamefont {{Blazek}}, \citenamefont {{Bocquet}}, \citenamefont {{Brooks}}, \citenamefont
  {{Burke}}, \citenamefont {{Carnero Rosell}}, \citenamefont {{Carretero}}, \citenamefont {{Cawthon}}, \citenamefont {{da Costa}}, \citenamefont {{da Silva Pereira}}, \citenamefont {{Diehl}}, \citenamefont {{Dodelson}}, \citenamefont {{Doel}}, \citenamefont {{Everett}}, \citenamefont {{Frohmaier}}, \citenamefont {{Garc{\'\i}a-Bellido}}, \citenamefont {{Gruen}}, \citenamefont {{Gutierrez}}, \citenamefont {{Herner}}, \citenamefont {{Hinton}}, \citenamefont {{Hollowood}}, \citenamefont {{Honscheid}}, \citenamefont {{Huterer}}, \citenamefont {{James}}, \citenamefont {{Jeffrey}}, \citenamefont {{Kuehn}}, \citenamefont {{Lahav}}, \citenamefont {{Lee}}, \citenamefont {{Lidman}}, \citenamefont {{Marshall}}, \citenamefont {{Mena-Fern{\'a}ndez}}, \citenamefont {{Menanteau}}, \citenamefont {{Miquel}}, \citenamefont {{Muir}}, \citenamefont {{Myles}}, \citenamefont {{Ogando}}, \citenamefont {{Paterno}}, \citenamefont {{Plazas Malag{\'o}n}}, \citenamefont {{Porredon}}, \citenamefont {{Prat}}, \citenamefont {{Nichol}},
  \citenamefont {{Romer}}, \citenamefont {{Roodman}}, \citenamefont {{Sanchez}}, \citenamefont {{Sanchez Cid}}, \citenamefont {{Sevilla-Noarbe}}, \citenamefont {{Suchyta}}, \citenamefont {{Swanson}}, \citenamefont {{To}}, \citenamefont {{Tucker}}, \citenamefont {{Walker}}, \citenamefont {{Weaverdyck}},\ and\ \citenamefont {{Aguena}}}]{25Dovekie_DESY5SNe}%
  \BibitemOpen
  \bibfield  {author} {\bibinfo {author} {\bibfnamefont {B.}~\bibnamefont {{Popovic}}}, \bibinfo {author} {\bibfnamefont {P.}~\bibnamefont {{Shah}}}, \bibinfo {author} {\bibfnamefont {W.~D.}\ \bibnamefont {{Kenworthy}}}, \emph {et~al.},\ }\href {https://doi.org/10.1093/mnras/stag632} {\bibfield  {journal} {\bibinfo  {journal} {\mnras}\ }\textbf {\bibinfo {volume} {548}},\ \bibinfo {pages} {stag632} (\bibinfo {year} {2026})},\ \Eprint {https://arxiv.org/abs/2511.07517} {arXiv:2511.07517 [astro-ph.CO]} \BibitemShut {NoStop}%
\bibitem [{\citenamefont {Lewis}\ \emph {et~al.}(2000)\citenamefont {Lewis}, \citenamefont {Challinor},\ and\ \citenamefont {Lasenby}}]{00Lewis_CAMB}%
  \BibitemOpen
  \bibfield  {author} {\bibinfo {author} {\bibfnamefont {A.}~\bibnamefont {Lewis}}, \bibinfo {author} {\bibfnamefont {A.}~\bibnamefont {Challinor}},\ and\ \bibinfo {author} {\bibfnamefont {A.}~\bibnamefont {Lasenby}},\ }\href {https://doi.org/10.1086/309179} {\bibfield  {journal} {\bibinfo  {journal} {\apj}\ }\textbf {\bibinfo {volume} {538}},\ \bibinfo {pages} {473} (\bibinfo {year} {2000})},\ \Eprint {https://arxiv.org/abs/astro-ph/9911177} {arXiv:astro-ph/9911177} \BibitemShut {NoStop}%
\bibitem [{\citenamefont {{Howlett}}\ \emph {et~al.}(2012)\citenamefont {{Howlett}}, \citenamefont {{Lewis}}, \citenamefont {{Hall}},\ and\ \citenamefont {{Challinor}}}]{12Howlett_CAMB}%
  \BibitemOpen
  \bibfield  {author} {\bibinfo {author} {\bibfnamefont {C.}~\bibnamefont {{Howlett}}}, \bibinfo {author} {\bibfnamefont {A.}~\bibnamefont {{Lewis}}}, \bibinfo {author} {\bibfnamefont {A.}~\bibnamefont {{Hall}}},\ and\ \bibinfo {author} {\bibfnamefont {A.}~\bibnamefont {{Challinor}}},\ }\href {https://doi.org/10.1088/1475-7516/2012/04/027} {\bibfield  {journal} {\bibinfo  {journal} {\jcap}\ }\textbf {\bibinfo {volume} {2012}},\ \bibinfo {eid} {027} (\bibinfo {year} {2012})},\ \Eprint {https://arxiv.org/abs/1201.3654} {arXiv:1201.3654 [astro-ph.CO]} \BibitemShut {NoStop}%
\bibitem [{\citenamefont {{Roy Choudhury}}\ and\ \citenamefont {{Okumura}}(2024)}]{24Choudhury_cmb_bao}%
  \BibitemOpen
  \bibfield  {author} {\bibinfo {author} {\bibfnamefont {S.}~\bibnamefont {{Roy Choudhury}}}\ and\ \bibinfo {author} {\bibfnamefont {T.}~\bibnamefont {{Okumura}}},\ }\href {https://doi.org/10.3847/2041-8213/ad8c26} {\bibfield  {journal} {\bibinfo  {journal} {\apjl}\ }\textbf {\bibinfo {volume} {976}},\ \bibinfo {eid} {L11} (\bibinfo {year} {2024})},\ \Eprint {https://arxiv.org/abs/2409.13022} {arXiv:2409.13022 [astro-ph.CO]} \BibitemShut {NoStop}%
\bibitem [{\citenamefont {{Ye}}\ and\ \citenamefont {{Lin}}(2025)}]{25Ye_cmb_bao}%
  \BibitemOpen
  \bibfield  {author} {\bibinfo {author} {\bibfnamefont {G.}~\bibnamefont {{Ye}}}\ and\ \bibinfo {author} {\bibfnamefont {S.-J.}\ \bibnamefont {{Lin}}},\ }\href {https://doi.org/10.48550/arXiv.2505.02207} {\bibfield  {journal} {\bibinfo  {journal} {arXiv e-prints}\ ,\ \bibinfo {eid} {arXiv:2505.02207}} (\bibinfo {year} {2025})},\ \Eprint {https://arxiv.org/abs/2505.02207} {arXiv:2505.02207 [astro-ph.CO]} \BibitemShut {NoStop}%
\bibitem [{\citenamefont {{Ong}}\ \emph {et~al.}(2025)\citenamefont {{Ong}}, \citenamefont {{Yallup}},\ and\ \citenamefont {{Handley}}}]{25Ong_cmb_bao}%
  \BibitemOpen
  \bibfield  {author} {\bibinfo {author} {\bibfnamefont {D.~D.~Y.}\ \bibnamefont {{Ong}}}, \bibinfo {author} {\bibfnamefont {D.}~\bibnamefont {{Yallup}}},\ and\ \bibinfo {author} {\bibfnamefont {W.}~\bibnamefont {{Handley}}},\ }\href {https://doi.org/10.48550/arXiv.2511.10631} {\bibfield  {journal} {\bibinfo  {journal} {arXiv e-prints}\ ,\ \bibinfo {eid} {arXiv:2511.10631}} (\bibinfo {year} {2025})},\ \Eprint {https://arxiv.org/abs/2511.10631} {arXiv:2511.10631 [astro-ph.CO]} \BibitemShut {NoStop}%
\bibitem [{\citenamefont {{Gu}}\ \emph {et~al.}(2025)\citenamefont {{Gu}}, \citenamefont {{Wang}}, \citenamefont {{Wang}}, \citenamefont {{Zhao}}, \citenamefont {{Pogosian}}, \citenamefont {{Koyama}}, \citenamefont {{Peacock}}, \citenamefont {{Cai}}, \citenamefont {{Cervantes-Cota}}, \citenamefont {{Ishak}}, \citenamefont {{Shafieloo}}, \citenamefont {{Zhao}}, \citenamefont {{Ahlen}}, \citenamefont {{Bianchi}}, \citenamefont {{Brooks}}, \citenamefont {{Claybaugh}}, \citenamefont {{Cole}}, \citenamefont {{de la Macorra}}, \citenamefont {{de Mattia}}, \citenamefont {{Doel}}, \citenamefont {{Ferraro}}, \citenamefont {{Forero-Romero}}, \citenamefont {{Gazta{\~n}aga}}, \citenamefont {{Gontcho A Gontcho}}, \citenamefont {{Gutierrez}}, \citenamefont {{Hahn}}, \citenamefont {{Howlett}}, \citenamefont {{Kehoe}}, \citenamefont {{Kirkby}}, \citenamefont {{Kneib}}, \citenamefont {{Kremin}}, \citenamefont {{Lahav}}, \citenamefont {{Landriau}}, \citenamefont {{Le Guillou}}, \citenamefont {{Leauthaud}}, \citenamefont
  {{Levi}}, \citenamefont {{Manera}}, \citenamefont {{Meisner}}, \citenamefont {{Miquel}}, \citenamefont {{Moustakas}}, \citenamefont {{Mu{\~n}oz-Guti{\'e}rrez}}, \citenamefont {{Nadathur}}, \citenamefont {{Newman}}, \citenamefont {{Palanque-Delabrouille}}, \citenamefont {{Percival}}, \citenamefont {{Prada}}, \citenamefont {{P{\'e}rez-R{\`a}fols}}, \citenamefont {{Rossi}}, \citenamefont {{Samushia}}, \citenamefont {{Sanchez}}, \citenamefont {{Schlegel}}, \citenamefont {{Seo}}, \citenamefont {{Sprayberry}}, \citenamefont {{Tarl{\'e}}}, \citenamefont {{Walther}}, \citenamefont {{Weaver}}, \citenamefont {{Zarrouk}}, \citenamefont {{Zhao}}, \citenamefont {{Zhou}},\ and\ \citenamefont {{Zou}}}]{25Gu_cmb_bao}%
  \BibitemOpen
  \bibfield  {author} {\bibinfo {author} {\bibfnamefont {G.}~\bibnamefont {{Gu}}}, \bibinfo {author} {\bibfnamefont {X.}~\bibnamefont {{Wang}}}, \bibinfo {author} {\bibfnamefont {Y.}~\bibnamefont {{Wang}}}, \emph {et~al.},\ }\href {https://doi.org/10.1038/s41550-025-02669-6} {\bibfield  {journal} {\bibinfo  {journal} {Nature Astronomy}\ }\textbf {\bibinfo {volume} {9}},\ \bibinfo {pages} {1879} (\bibinfo {year} {2025})},\ \Eprint {https://arxiv.org/abs/2504.06118} {arXiv:2504.06118 [astro-ph.CO]} \BibitemShut {NoStop}%
\bibitem [{\citenamefont {{Ferreira}}\ \emph {et~al.}(2026)\citenamefont {{Ferreira}}, \citenamefont {{McDonough}}, \citenamefont {{Balkenhol}}, \citenamefont {{Kallosh}}, \citenamefont {{Knox}},\ and\ \citenamefont {{Linde}}}]{26Ferreira_cmb_bao}%
  \BibitemOpen
  \bibfield  {author} {\bibinfo {author} {\bibfnamefont {E.~G.~M.}\ \bibnamefont {{Ferreira}}}, \bibinfo {author} {\bibfnamefont {E.}~\bibnamefont {{McDonough}}}, \bibinfo {author} {\bibfnamefont {L.}~\bibnamefont {{Balkenhol}}}, \emph {et~al.},\ }\href {https://doi.org/10.1103/lq71-b84v} {\bibfield  {journal} {\bibinfo  {journal} {\prd}\ }\textbf {\bibinfo {volume} {113}},\ \bibinfo {eid} {043524} (\bibinfo {year} {2026})},\ \Eprint {https://arxiv.org/abs/2507.12459} {arXiv:2507.12459 [astro-ph.CO]} \BibitemShut {NoStop}%
\bibitem [{\citenamefont {{Lee}}(2026)}]{26Lee_cmb_bao}%
  \BibitemOpen
  \bibfield  {author} {\bibinfo {author} {\bibfnamefont {S.}~\bibnamefont {{Lee}}},\ }\href {https://doi.org/10.1016/j.aop.2026.170453} {\bibfield  {journal} {\bibinfo  {journal} {Annals of Physics}\ }\textbf {\bibinfo {volume} {489}},\ \bibinfo {eid} {170453} (\bibinfo {year} {2026})},\ \Eprint {https://arxiv.org/abs/2507.01380} {arXiv:2507.01380 [astro-ph.CO]} \BibitemShut {NoStop}%
\bibitem [{\citenamefont {Hergt}(2026)}]{Zenodo2026_hergt_glitch}%
  \BibitemOpen
  \bibfield  {author} {\bibinfo {author} {\bibfnamefont {L.~T.}\ \bibnamefont {Hergt}},\ }\href {https://doi.org/10.5281/zenodo.23063691} {10.5281/zenodo.23063691} (\bibinfo {year} {2026}),\ \bibinfo {note} {\textit{Inference products and plotting code for ``Is DESI Seeing Dynamical Dark Energy, or a Cosmic Glitch in Gravity?''}}\BibitemShut {NoStop}%
\bibitem [{\citenamefont {{Doran}}\ and\ \citenamefont {{Robbers}}(2006)}]{06DoranEDE}%
  \BibitemOpen
  \bibfield  {author} {\bibinfo {author} {\bibfnamefont {M.}~\bibnamefont {{Doran}}}\ and\ \bibinfo {author} {\bibfnamefont {G.}~\bibnamefont {{Robbers}}},\ }\href {https://doi.org/10.1088/1475-7516/2006/06/026} {\bibfield  {journal} {\bibinfo  {journal} {\jcap}\ }\textbf {\bibinfo {volume} {2006}},\ \bibinfo {eid} {026} (\bibinfo {year} {2006})},\ \Eprint {https://arxiv.org/abs/astro-ph/0601544} {arXiv:astro-ph/0601544 [astro-ph]} \BibitemShut {NoStop}%
\bibitem [{\citenamefont {{Akarsu}}\ \emph {et~al.}(2020)\citenamefont {{Akarsu}}, \citenamefont {{Barrow}}, \citenamefont {{Escamilla}},\ and\ \citenamefont {{Vazquez}}}]{20Akarsu_DE_signswitch}%
  \BibitemOpen
  \bibfield  {author} {\bibinfo {author} {\bibfnamefont {{\"O}.}~\bibnamefont {{Akarsu}}}, \bibinfo {author} {\bibfnamefont {J.~D.}\ \bibnamefont {{Barrow}}}, \bibinfo {author} {\bibfnamefont {L.~A.}\ \bibnamefont {{Escamilla}}},\ and\ \bibinfo {author} {\bibfnamefont {J.~A.}\ \bibnamefont {{Vazquez}}},\ }\href {https://doi.org/10.1103/PhysRevD.101.063528} {\bibfield  {journal} {\bibinfo  {journal} {\prd}\ }\textbf {\bibinfo {volume} {101}},\ \bibinfo {eid} {063528} (\bibinfo {year} {2020})},\ \Eprint {https://arxiv.org/abs/1912.08751} {arXiv:1912.08751 [astro-ph.CO]} \BibitemShut {NoStop}%
\bibitem [{\citenamefont {{Akarsu}}\ \emph {et~al.}(2024)\citenamefont {{Akarsu}}, \citenamefont {{De Felice}}, \citenamefont {{Di Valentino}}, \citenamefont {{Kumar}}, \citenamefont {{Nunes}}, \citenamefont {{{\"O}z{\"u}lker}}, \citenamefont {{Vazquez}},\ and\ \citenamefont {{Yadav}}}]{24Akarsu_LsCDM_MMG}%
  \BibitemOpen
  \bibfield  {author} {\bibinfo {author} {\bibfnamefont {{\"O}.}~\bibnamefont {{Akarsu}}}, \bibinfo {author} {\bibfnamefont {A.}~\bibnamefont {{De Felice}}}, \bibinfo {author} {\bibfnamefont {E.}~\bibnamefont {{Di Valentino}}}, \emph {et~al.},\ }\href {https://doi.org/10.1103/PhysRevD.110.103527} {\bibfield  {journal} {\bibinfo  {journal} {\prd}\ }\textbf {\bibinfo {volume} {110}},\ \bibinfo {eid} {103527} (\bibinfo {year} {2024})},\ \Eprint {https://arxiv.org/abs/2406.07526} {arXiv:2406.07526 [astro-ph.CO]} \BibitemShut {NoStop}%
\bibitem [{\citenamefont {{G{\"o}k{\c{c}}en}}\ \emph {et~al.}(2026)\citenamefont {{G{\"o}k{\c{c}}en}}, \citenamefont {{Akarsu}},\ and\ \citenamefont {{Di Valentino}}}]{26Gokcen_NECB}%
  \BibitemOpen
  \bibfield  {author} {\bibinfo {author} {\bibfnamefont {M.}~\bibnamefont {{G{\"o}k{\c{c}}en}}}, \bibinfo {author} {\bibfnamefont {{\"O}.}~\bibnamefont {{Akarsu}}},\ and\ \bibinfo {author} {\bibfnamefont {E.}~\bibnamefont {{Di Valentino}}},\ }\href {https://doi.org/10.1016/j.dark.2026.102273} {\bibfield  {journal} {\bibinfo  {journal} {Physics of the Dark Universe}\ }\textbf {\bibinfo {volume} {52}},\ \bibinfo {eid} {102273} (\bibinfo {year} {2026})},\ \Eprint {https://arxiv.org/abs/2602.21169} {arXiv:2602.21169 [astro-ph.CO]} \BibitemShut {NoStop}%
\bibitem [{\citenamefont {{Hu}}(2008)}]{08HuPPFDE}%
  \BibitemOpen
  \bibfield  {author} {\bibinfo {author} {\bibfnamefont {W.}~\bibnamefont {{Hu}}},\ }\href {https://doi.org/10.1103/PhysRevD.77.103524} {\bibfield  {journal} {\bibinfo  {journal} {\prd}\ }\textbf {\bibinfo {volume} {77}},\ \bibinfo {eid} {103524} (\bibinfo {year} {2008})},\ \Eprint {https://arxiv.org/abs/0801.2433} {arXiv:0801.2433 [astro-ph]} \BibitemShut {NoStop}%
\bibitem [{\citenamefont {{Fang}}\ \emph {et~al.}(2008)\citenamefont {{Fang}}, \citenamefont {{Hu}},\ and\ \citenamefont {{Lewis}}}]{08FangPPFDE}%
  \BibitemOpen
  \bibfield  {author} {\bibinfo {author} {\bibfnamefont {W.}~\bibnamefont {{Fang}}}, \bibinfo {author} {\bibfnamefont {W.}~\bibnamefont {{Hu}}},\ and\ \bibinfo {author} {\bibfnamefont {A.}~\bibnamefont {{Lewis}}},\ }\href {https://doi.org/10.1103/PhysRevD.78.087303} {\bibfield  {journal} {\bibinfo  {journal} {\prd}\ }\textbf {\bibinfo {volume} {78}},\ \bibinfo {eid} {087303} (\bibinfo {year} {2008})},\ \Eprint {https://arxiv.org/abs/0808.3125} {arXiv:0808.3125 [astro-ph]} \BibitemShut {NoStop}%
\bibitem [{\citenamefont {{Torrado}}\ and\ \citenamefont {{Lewis}}(2021)}]{20Torrado_Cobaya}%
  \BibitemOpen
  \bibfield  {author} {\bibinfo {author} {\bibfnamefont {J.}~\bibnamefont {{Torrado}}}\ and\ \bibinfo {author} {\bibfnamefont {A.}~\bibnamefont {{Lewis}}},\ }\href {https://doi.org/10.1088/1475-7516/2021/05/057} {\bibfield  {journal} {\bibinfo  {journal} {\jcap}\ }\textbf {\bibinfo {volume} {2021}},\ \bibinfo {eid} {057} (\bibinfo {year} {2021})},\ \Eprint {https://arxiv.org/abs/2005.05290} {arXiv:2005.05290 [astro-ph.IM]} \BibitemShut {NoStop}%
\bibitem [{\citenamefont {{Gelman}}\ and\ \citenamefont {{Rubin}}(1992)}]{92Gelman}%
  \BibitemOpen
  \bibfield  {author} {\bibinfo {author} {\bibfnamefont {A.}~\bibnamefont {{Gelman}}}\ and\ \bibinfo {author} {\bibfnamefont {D.~B.}\ \bibnamefont {{Rubin}}},\ }\href {https://doi.org/10.1214/ss/1177011136} {\bibfield  {journal} {\bibinfo  {journal} {Statistical Science}\ }\textbf {\bibinfo {volume} {7}},\ \bibinfo {pages} {457} (\bibinfo {year} {1992})}\BibitemShut {NoStop}%
\bibitem [{\citenamefont {Handley}(2019)}]{19Anesthetic}%
  \BibitemOpen
  \bibfield  {author} {\bibinfo {author} {\bibfnamefont {W.}~\bibnamefont {Handley}},\ }\href {https://doi.org/10.21105/joss.01414} {\bibfield  {journal} {\bibinfo  {journal} {The Journal of Open Source Software}\ }\textbf {\bibinfo {volume} {4}},\ \bibinfo {pages} {1414} (\bibinfo {year} {2019})}\BibitemShut {NoStop}%
\bibitem [{\citenamefont {{Handley}}\ \emph {et~al.}(2015{\natexlab{a}})\citenamefont {{Handley}}, \citenamefont {{Hobson}},\ and\ \citenamefont {{Lasenby}}}]{15Handley_Polychord_stat}%
  \BibitemOpen
  \bibfield  {author} {\bibinfo {author} {\bibfnamefont {W.~J.}\ \bibnamefont {{Handley}}}, \bibinfo {author} {\bibfnamefont {M.~P.}\ \bibnamefont {{Hobson}}},\ and\ \bibinfo {author} {\bibfnamefont {A.~N.}\ \bibnamefont {{Lasenby}}},\ }\href {https://doi.org/10.1093/mnras/stv1911} {\bibfield  {journal} {\bibinfo  {journal} {\mnras}\ }\textbf {\bibinfo {volume} {453}},\ \bibinfo {pages} {4384} (\bibinfo {year} {2015}{\natexlab{a}})},\ \Eprint {https://arxiv.org/abs/1506.00171} {arXiv:1506.00171 [astro-ph.IM]} \BibitemShut {NoStop}%
\bibitem [{\citenamefont {{Handley}}\ \emph {et~al.}(2015{\natexlab{b}})\citenamefont {{Handley}}, \citenamefont {{Hobson}},\ and\ \citenamefont {{Lasenby}}}]{15Handley_polychord_cosmo}%
  \BibitemOpen
  \bibfield  {author} {\bibinfo {author} {\bibfnamefont {W.~J.}\ \bibnamefont {{Handley}}}, \bibinfo {author} {\bibfnamefont {M.~P.}\ \bibnamefont {{Hobson}}},\ and\ \bibinfo {author} {\bibfnamefont {A.~N.}\ \bibnamefont {{Lasenby}}},\ }\href {https://doi.org/10.1093/mnrasl/slv047} {\bibfield  {journal} {\bibinfo  {journal} {\mnras}\ }\textbf {\bibinfo {volume} {450}},\ \bibinfo {pages} {L61} (\bibinfo {year} {2015}{\natexlab{b}})},\ \Eprint {https://arxiv.org/abs/1502.01856} {arXiv:1502.01856 [astro-ph.CO]} \BibitemShut {NoStop}%
\bibitem [{\citenamefont {{Planck Collaboration}}(2020{\natexlab{b}})}]{18Plancklikelihood}%
  \BibitemOpen
  \bibfield  {author} {\bibinfo {author} {\bibnamefont {{Planck Collaboration}}},\ }\href {https://doi.org/10.1051/0004-6361/201936386} {\bibfield  {journal} {\bibinfo  {journal} {\aap}\ }\textbf {\bibinfo {volume} {641}},\ \bibinfo {pages} {A5} (\bibinfo {year} {2020}{\natexlab{b}})},\ \Eprint {https://arxiv.org/abs/1907.12875} {arXiv:1907.12875} \BibitemShut {NoStop}%
\bibitem [{\citenamefont {{Tristram}}\ \emph {et~al.}(2021)\citenamefont {{Tristram}}, \citenamefont {{Banday}}, \citenamefont {{G{\'o}rski}}, \citenamefont {{Keskitalo}}, \citenamefont {{Lawrence}}, \citenamefont {{Andersen}}, \citenamefont {{Barreiro}}, \citenamefont {{Borrill}}, \citenamefont {{Eriksen}}, \citenamefont {{Fernandez-Cobos}}, \citenamefont {{Kisner}}, \citenamefont {{Mart{\'\i}nez-Gonz{\'a}lez}}, \citenamefont {{Partridge}}, \citenamefont {{Scott}}, \citenamefont {{Svalheim}}, \citenamefont {{Thommesen}},\ and\ \citenamefont {{Wehus}}}]{21Tristram_Lollipop}%
  \BibitemOpen
  \bibfield  {author} {\bibinfo {author} {\bibfnamefont {M.}~\bibnamefont {{Tristram}}}, \bibinfo {author} {\bibfnamefont {A.~J.}\ \bibnamefont {{Banday}}}, \bibinfo {author} {\bibfnamefont {K.~M.}\ \bibnamefont {{G{\'o}rski}}}, \emph {et~al.},\ }\href {https://doi.org/10.1051/0004-6361/202039585} {\bibfield  {journal} {\bibinfo  {journal} {\aap}\ }\textbf {\bibinfo {volume} {647}},\ \bibinfo {eid} {A128} (\bibinfo {year} {2021})},\ \Eprint {https://arxiv.org/abs/2010.01139} {arXiv:2010.01139 [astro-ph.CO]} \BibitemShut {NoStop}%
\bibitem [{\citenamefont {{Carron}}\ \emph {et~al.}(2022)\citenamefont {{Carron}}, \citenamefont {{Mirmelstein}},\ and\ \citenamefont {{Lewis}}}]{22Carron_PR4lensing}%
  \BibitemOpen
  \bibfield  {author} {\bibinfo {author} {\bibfnamefont {J.}~\bibnamefont {{Carron}}}, \bibinfo {author} {\bibfnamefont {M.}~\bibnamefont {{Mirmelstein}}},\ and\ \bibinfo {author} {\bibfnamefont {A.}~\bibnamefont {{Lewis}}},\ }\href {https://doi.org/10.1088/1475-7516/2022/09/039} {\bibfield  {journal} {\bibinfo  {journal} {\jcap}\ }\textbf {\bibinfo {volume} {2022}},\ \bibinfo {eid} {039} (\bibinfo {year} {2022})},\ \Eprint {https://arxiv.org/abs/2206.07773} {arXiv:2206.07773 [astro-ph.CO]} \BibitemShut {NoStop}%
\bibitem [{\citenamefont {{Madhavacheril}}\ \emph {et~al.}(2024)\citenamefont {{Madhavacheril}}, \citenamefont {{Qu}}, \citenamefont {{Sherwin}}, \citenamefont {{MacCrann}}, \citenamefont {{Li}},\ and\ \citenamefont {{et al.}}}]{24Madhavacheril_ACTDR6lensing}%
  \BibitemOpen
  \bibfield  {author} {\bibinfo {author} {\bibfnamefont {M.~S.}\ \bibnamefont {{Madhavacheril}}}, \bibinfo {author} {\bibfnamefont {F.~J.}\ \bibnamefont {{Qu}}}, \bibinfo {author} {\bibfnamefont {B.~D.}\ \bibnamefont {{Sherwin}}}, \emph {et~al.},\ }\href {https://doi.org/10.3847/1538-4357/acff5f} {\bibfield  {journal} {\bibinfo  {journal} {\apj}\ }\textbf {\bibinfo {volume} {962}},\ \bibinfo {eid} {113} (\bibinfo {year} {2024})},\ \Eprint {https://arxiv.org/abs/2304.05203} {arXiv:2304.05203 [astro-ph.CO]} \BibitemShut {NoStop}%
\bibitem [{\citenamefont {{Qu}}\ \emph {et~al.}(2024)\citenamefont {{Qu}}, \citenamefont {{Sherwin}}, \citenamefont {{Madhavacheril}}, \citenamefont {{Han}}, \citenamefont {{Crowley}},\ and\ \citenamefont {{et al.}}}]{24Qu_ACTDR6lensing}%
  \BibitemOpen
  \bibfield  {author} {\bibinfo {author} {\bibfnamefont {F.~J.}\ \bibnamefont {{Qu}}}, \bibinfo {author} {\bibfnamefont {B.~D.}\ \bibnamefont {{Sherwin}}}, \bibinfo {author} {\bibfnamefont {M.~S.}\ \bibnamefont {{Madhavacheril}}}, \emph {et~al.},\ }\href {https://doi.org/10.3847/1538-4357/acfe06} {\bibfield  {journal} {\bibinfo  {journal} {\apj}\ }\textbf {\bibinfo {volume} {962}},\ \bibinfo {eid} {112} (\bibinfo {year} {2024})},\ \Eprint {https://arxiv.org/abs/2304.05202} {arXiv:2304.05202 [astro-ph.CO]} \BibitemShut {NoStop}%
\bibitem [{\citenamefont {Scolnic}\ \emph {et~al.}(2022)\citenamefont {Scolnic} \emph {et~al.}}]{21Scolnic_PantheonP}%
  \BibitemOpen
  \bibfield  {author} {\bibinfo {author} {\bibfnamefont {D.}~\bibnamefont {Scolnic}} \emph {et~al.},\ }\href {https://doi.org/10.3847/1538-4357/ac8b7a} {\bibfield  {journal} {\bibinfo  {journal} {Astrophys. J.}\ }\textbf {\bibinfo {volume} {938}},\ \bibinfo {pages} {113} (\bibinfo {year} {2022})},\ \Eprint {https://arxiv.org/abs/2112.03863} {arXiv:2112.03863 [astro-ph.CO]} \BibitemShut {NoStop}%
\bibitem [{\citenamefont {{Rubin}}\ \emph {et~al.}(2025)\citenamefont {{Rubin}}, \citenamefont {{Aldering}}, \citenamefont {{Betoule}}, \citenamefont {{Fruchter}}, \citenamefont {{Huang}}, \citenamefont {{Kim}}, \citenamefont {{Lidman}}, \citenamefont {{Linder}}, \citenamefont {{Perlmutter}}, \citenamefont {{Ruiz-Lapuente}},\ and\ \citenamefont {{Suzuki}}}]{25Rubin_UnionY3}%
  \BibitemOpen
  \bibfield  {author} {\bibinfo {author} {\bibfnamefont {D.}~\bibnamefont {{Rubin}}}, \bibinfo {author} {\bibfnamefont {G.}~\bibnamefont {{Aldering}}}, \bibinfo {author} {\bibfnamefont {M.}~\bibnamefont {{Betoule}}}, \emph {et~al.},\ }\href {https://doi.org/10.3847/1538-4357/adc0a5} {\bibfield  {journal} {\bibinfo  {journal} {\apj}\ }\textbf {\bibinfo {volume} {986}},\ \bibinfo {eid} {231} (\bibinfo {year} {2025})},\ \Eprint {https://arxiv.org/abs/2311.12098} {arXiv:2311.12098 [astro-ph.CO]} \BibitemShut {NoStop}%
\end{thebibliography}%

\clearpage
\onecolumngrid
\appendix
\section*{Supplemental Material}
\refstepcounter{smref}
\label{app:supplement}
\setcounter{figure}{0}
\setcounter{table}{0}
\setcounter{equation}{0}
\renewcommand{\thefigure}{S\arabic{figure}}
\renewcommand{\thetable}{S\arabic{table}}
\renewcommand{\theequation}{S\arabic{equation}}

\subsection{Model and perturbation prescription}

The CGG model modifies \LCDM by allowing the gravitational constant entering the cosmological Friedmann equation to differ from the locally measured Newtonian value, $\Og \equiv 1 - \GN/\Gc$. In the effective-fluid implementation used for our numerical analysis, the dark-energy sector contains a constant vacuum density plus a component equal to a fixed fraction \Og of the critical density,
\begin{equation}
    \rho_\mathrm{DE} = \rho_\Lambda + \Og\, \rho_\mathrm{crit}.
\end{equation}
Since $\rho_\mathrm{crit}$ itself contains $\rho_\mathrm{DE}$, this definition is implicit. Solving for the effective dark-energy density gives
\begin{equation}
    \rho_\mathrm{DE}
    =
    \frac{\Og \rho_\mathrm{nonDE} + \rho_\Lambda}{1-\Og},
    \label{eq:sm_rho_CGG}
\end{equation}
where $\rho_\mathrm{nonDE}=\rho_\mathrm{m}+\rho_\mathrm{r}+\rho_\nu$. The associated effective equation of state follows from energy conservation,
\begin{equation}
    1+w_\mathrm{DE}
    =
    \frac{\Og}{(1-\Og)\rho_\mathrm{DE}}
    \sum_i \rho_i(1+w_i),
    \label{eq:sm_w_CGG}
\end{equation}
with $w_i$ the equation-of-state parameter of non-dark-energy species $i$. The effective component tracks the dominant species: $w_\mathrm{DE}=1/3$ during radiation domination, $w_\mathrm{DE}=0$ during matter domination, and $w_\mathrm{DE}\simeq -1$ once the constant term dominates.

The two signs of \Og have different effective-fluid interpretations. For $\Og>0$, the effective dark-energy density remains positive and resembles a tracking early-dark-energy component~\cite{06DoranEDE}. For $\Og<0$, the effective density is negative at early times and changes sign before becoming positive today; the equation-of-state variable then diverges at the zero crossing. In the CGG interpretation this is not a physical instability; it is a limitation of describing a modified-gravity mismatch as a sign-changing effective fluid, analogous to other sign-switching dark-energy parameterizations~\cite{20Akarsu_DE_signswitch,24Akarsu_LsCDM_MMG,26Gokcen_NECB}.

We treat perturbations using the parameterized post-Friedmann (PPF) prescription~\cite{08HuPPFDE,08FangPPFDE}, as implemented in the PPF dark-energy module of the modified \texttt{CAMB} code used in Refs.~\cite{24WenCGG,24WenCGG_essay}. The fiducial choice is the standard rest-frame sound speed $\cs=1$, i.e., dark-energy perturbations that propagate at the speed of light, as for a canonical scalar field. Because cuscuton-like realizations are smooth and non-dynamical, corresponding formally to $\cs\to\infty$, we also test $\cs^2=1000$ as a numerical proxy for this limit, and a run in which $\log_{10}\cs^2$ is varied over a broad prior. Values $\cs>1$ correspond to superluminal propagation for an ordinary fluid, but not acausal propagation in a theory with a preferred foliation, where the khronon defines the global time coordinate; in practice, Ref.~\cite{Robbers:2007ca} shows that models become effectively indistinguishable for $\cs^2 \gtrsim 10$, once the dark-energy Jeans scale exceeds the observable range. These sound-speed tests are summarized in the sound-speed section below. At the level of a fluid description, PPF and an exact extended-Ho\v{r}ava perturbation treatment should agree once the background stress tensor and effective sound speed are matched; this equivalence is explicit in the smooth $\cs\to\infty$ cuscuton/minimal-Ho\v{r}ava limit~\cite{07CuscutonCosmo,09CuscutonHovrava}, while for finite \cs we use PPF phenomenologically.

The data constrain \Og through both background and perturbation effects. At the background level, the modified expansion history changes the BAO and SN distances and the angular projection of the CMB acoustic scale (note the shift in $100\,\theta_\ast$ between \LCDM and CGG in \cref{tab:constraints}); at the perturbation level, the altered evolution of the gravitational potentials modifies the integrated Sachs--Wolfe contribution at low multipoles and the CMB lensing amplitude, as well as the growth of structure discussed next.

The same background deformation arises naturally in Lorentz-violating theories with a preferred foliation, including Ho\v{r}ava--Lifshitz gravity~\cite{09Hovrava}, Einstein-aether theory~\cite{10Jacobson_Aether}, and the cuscuton representation~\cite{07CuscutonCosmo,09CuscutonHovrava}. Restricted or minimal Ho\v{r}ava gravity provides a particularly close theoretical target~\cite{10Pons_minimal_Horava,25JacobsonPulakkat_mHg}. The main text therefore describes the result as empirical evidence for the cosmological--Newtonian gravitational mismatch realized by these theories, not as a unique detection of any one complete theory.

\begin{table*}[t]
    \centering
    \caption{\label{tab:constraints}
        Marginalized means and \SI{68}{\percent} credible intervals for \LCDM, CGG, and \wwCDM, for CMB+BAO (\Hillik + DESI~DR2) and for CMB+L+BAO+SN (further adding \Planck+ACT+SPT lensing and DES~Dovekie SN). Parameters in the upper part of the table are sampled; those in the lower part are derived. $H_0$ is in $\mathrm{km\,s^{-1}\,Mpc^{-1}}$ and $r_\mathrm{d}$ in Mpc.
    }
\begin{tabularx}{\textwidth}{ l r@{${}\pm{}$}l r@{${}\pm{}$}l r@{${}\pm{}$}l r@{${}\pm{}$}l r@{${}\pm{}$}l r@{${}\pm{}$}l }
\toprule
                                & \multicolumn{6}{c}{CMB + BAO} & \multicolumn{6}{c}{CMB + L + BAO + SN} \\
                                 \cmidrule{3-6}                  \cmidrule{9-12}
                                & \multicolumn{2}{c}{$\Lambda$CDM} & \multicolumn{2}{c}{$\mathrm{CGG}$} & \multicolumn{2}{c}{$w_0w_a$CDM} & \multicolumn{2}{c}{$\Lambda$CDM} & \multicolumn{2}{c}{$\mathrm{CGG}$} & \multicolumn{2}{c}{$w_0w_a$CDM} \\
\midrule
    $\Omega_\mathrm{g}$         & \multicolumn{2}{c}{$0$}          &       $-0.0081$ & $0.0025$         & \multicolumn{2}{c}{$0$}         & \multicolumn{2}{c}{$0$}          &       $-0.0073$ & $0.0022$         & \multicolumn{2}{c}{$0$}         \\
\addlinespace[\medskipamount]
    $w_0$                       & \multicolumn{2}{c}{$-1$}         & \multicolumn{2}{c}{$-1$}           &        $-0.44$ & $0.20$         & \multicolumn{2}{c}{$-1$}         & \multicolumn{2}{c}{$-1$}           &        $-0.80$ & $0.06$         \\
    $w_a$                       & \multicolumn{2}{c}{$0$}          & \multicolumn{2}{c}{$0$}            &         $-1.7$ & $0.6$          & \multicolumn{2}{c}{$0$}          & \multicolumn{2}{c}{$0$}            &        $-0.75$ & $0.21$         \\
\addlinespace[\medskipamount]
    $100\,\Omega_\mathrm{b}h^2$ &        $2.237$ & $0.008$         &         $2.231$ & $0.009$          &        $2.231$ & $0.009$        &        $2.235$ & $0.008$         &         $2.231$ & $0.009$          &        $2.232$ & $0.009$        \\
    $10\,\Omega_\mathrm{c}h^2$  &        $1.177$ & $0.006$         &         $1.169$ & $0.006$          &        $1.195$ & $0.008$        &        $1.181$ & $0.006$         &         $1.173$ & $0.006$          &        $1.193$ & $0.007$        \\
    $H_0$                       &        $68.21$ & $0.25$          &         $68.31$ & $0.26$           &         $63.8$ & $1.8$          &        $68.01$ & $0.24$          &         $68.19$ & $0.25$           &         $67.4$ & $0.5$          \\
    $\tau_\mathrm{reio}$        &        $0.066$ & $0.006$         &         $0.060$ & $0.006$          &        $0.062$ & $0.006$        &        $0.068$ & $0.006$         &         $0.060$ & $0.006$          &        $0.063$ & $0.006$        \\
    $\ln(10^{10}A_\mathrm{s})$  &        $3.064$ & $0.013$         &         $3.048$ & $0.014$          &        $3.054$ & $0.013$        &        $3.068$ & $0.010$         &         $3.045$ & $0.012$          &        $3.056$ & $0.011$        \\
    $n_\mathrm{s}$              &       $0.9713$ & $0.0030$        &        $0.9692$ & $0.0030$         &       $0.9673$ & $0.0031$       &       $0.9705$ & $0.0029$        &        $0.9689$ & $0.0030$         &       $0.9679$ & $0.0031$       \\
\midrule
    $\Omega_\Lambda$            &       $0.6975$ & $0.0034$        &         $0.708$ & $0.005$          &        $0.649$ & $0.021$        &       $0.6948$ & $0.0033$        &         $0.706$ & $0.005$          &        $0.687$ & $0.005$        \\
    $\Omega_\mathrm{m}$         &       $0.3024$ & $0.0034$        &        $0.2998$ & $0.0035$         &        $0.351$ & $0.021$        &       $0.3051$ & $0.0033$        &        $0.3016$ & $0.0034$         &        $0.313$ & $0.005$        \\
    $100\,\theta_\ast$          &      $1.04123$ & $0.00019$       &       $1.04182$ & $0.00027$        &      $1.04105$ & $0.00020$      &      $1.04121$ & $0.00019$       &       $1.04174$ & $0.00025$        &      $1.04108$ & $0.00019$      \\
    $\sigma_8$                  &        $0.813$ & $0.006$         &         $0.836$ & $0.009$          &        $0.788$ & $0.016$        &        $0.816$ & $0.004$         &         $0.832$ & $0.006$          &        $0.818$ & $0.007$        \\
    $S_8$                       &        $0.816$ & $0.008$         &         $0.835$ & $0.010$          &        $0.851$ & $0.013$        &        $0.823$ & $0.006$         &         $0.834$ & $0.007$          &        $0.836$ & $0.007$        \\
    $r_\mathrm{d}$              &       $147.72$ & $0.18$          &        $148.59$ & $0.32$           &       $147.29$ & $0.22$         &       $147.62$ & $0.17$          &        $148.44$ & $0.30$           &       $147.34$ & $0.20$         \\
    $r_\mathrm{d}h$             &        $100.8$ & $0.5$           &         $101.5$ & $0.5$            &         $94.0$ & $2.7$          &        $100.4$ & $0.4$           &         $101.2$ & $0.5$            &         $99.3$ & $0.8$          \\
\bottomrule
\end{tabularx}
\end{table*}

\subsection{Linear growth of structure}
\label{sec:sm_growth}

The background expansion in CGG is governed by \Gc, $H^2=8\pi\Gc\rho/3$, while the Poisson equation for sub-horizon perturbations retains the locally measured coupling, $\nabla^2\Phi=4\pi\GN\bar\rho\,\delta$. During matter domination the linear growth equation therefore reads
\begin{equation}
    \ddot\delta+2H\dot\delta = 4\pi\GN\bar\rho_\mathrm{m}\,\delta = \tfrac{3}{2}(1-\Og)H^2\delta .
\end{equation}
With $a\propto t^{2/3}$ the growing mode is $\delta\propto a^{p}$ with $p=\tfrac{3}{4}\bigl[-\tfrac13+\sqrt{\tfrac19+\tfrac83(1-\Og)}\bigr]$, i.e.,
\begin{equation}
    \delta \propto a^{\,1-3\Og/5+\mathcal{O}(\Og^2)}, \qquad \Phi\propto\delta/a\propto a^{-3\Og/5}.
    \label{eq:sm_growth}
\end{equation}
For $\Og<0$ structure grows slightly faster than in \LCDM and the potential grows slowly during matter domination, sourcing an early-time integrated Sachs--Wolfe effect, opposite in sign to the late-time ISW from dark-energy-driven decay. The corresponding growth rate is $f\equiv\mathrm{d}\ln\delta/\mathrm{d}\ln a\simeq1-3\Og/5$ in the matter era, a roughly $\SI{0.5}{\percent}$ enhancement for the best-fit \Og. The PPF implementation reproduces this behavior: for $\cs\gg1$ the effective fluid is smooth on sub-horizon scales and enters the growth equation only through $H(a)$, which is exactly the modification above (and for $\cs=1$ its perturbations are negligible inside the horizon). The dominant observable consequence is cumulative. Integrated from recombination to today, \cref{eq:sm_growth} enhances the linear growth factor relative to \LCDM by $(1+z_\ast)^{-3\Og/5}\simeq1.035$ for $\Og=-0.0081$, i.e., \SI{3.5}{\percent} extra; solving the full growth equation, including the onset of $\Lambda$ domination which ends the enhanced growth, gives \SI{3.1}{\percent} at fixed \Om and \SI{2.9}{\percent} with the slightly lower posterior \Om of CGG. The factor scales rapidly with the glitch, \SI{2.1}{\percent}, \SI{6.5}{\percent}, and \SI{8.8}{\percent} for $\Og=-0.005$, $-0.015$, and $-0.02$, but at fixed CMB data it is partly absorbed along the $A_\mathrm{s}$--$\tau$ degeneracy: the CGG posterior has lower $\ln(10^{10}A_\mathrm{s})$ ($3.048$ versus $3.064$) and lower $\tau_\mathrm{reio}$ ($0.060$ versus $0.066$), so that the combination $A_\mathrm{s}e^{-2\tau}$ (fixed by the acoustic peaks) is unchanged, and the net effect is a \SI{2.8}{\percent} higher $\sigma_8$ (\cref{tab:constraints}). CMB lensing, which measures the late-time amplitude directly, and an independent determination of $\tau$ break this degeneracy and therefore probe \Og through structure growth rather than through distances. For the redshifts probed by galaxy surveys the picture is different: solving the linear growth equation with the CMB+BAO posterior means of \cref{tab:constraints} gives $f\sigma_8$ higher than in \LCDM by \SI{2.7}{\percent} at $z=0.3$, rising to \SI{3.1}{\percent} at $z=1.5$, while $f$ itself differs by less than \SI{0.3}{\percent}, because the slightly lower \Om of CGG largely compensates the enhanced growth rate at late times. The redshift-space-distortion signature of the glitch is therefore an almost constant $\SI{3}{\percent}$ offset in the amplitude of $f\sigma_8(z)$.

\subsection{Data combinations and sampling}
\label{sec:sm_data}

Theoretical predictions are computed with the modified \texttt{CAMB} implementation of Refs.~\cite{24WenCGG,24WenCGG_essay}.\footnote{\url{https://github.com/lukashergt/CAMB/tree/CGG}} As listed in \cref{tab:priors}, we sample with \texttt{Cobaya}~\cite{20Torrado_Cobaya} over the six standard \LCDM parameters and the cosmic glitch parameter~\Og, with a single massive neutrino of mass $m_\nu=0.06\,\mathrm{eV}$ and $N_\mathrm{eff}=3.044$. Markov chains are run until the Gelman--Rubin statistic satisfies $R-1\leq0.01$~\cite{92Gelman}, and visualized with \texttt{anesthetic}~\cite{19Anesthetic}. Bayesian evidence is computed with \texttt{PolyChord}~\cite{15Handley_Polychord_stat,15Handley_polychord_cosmo}, with evidence decompositions produced using \texttt{anesthetic}~\cite{19Anesthetic}.
The PolyChord runs used the uniform priors $[-0.1,0.1]$ for $\Og$, $[-2,0]$ for $w_0$, and $[-3,2]$ for $w_a$, with an additional rejection prior from the condition $w_0+w_a<0$, as listed in \cref{tab:priors}.

\begin{table*}[tbp]
    \centering
    \caption{\label{tab:priors}
        Prior ranges of the cosmological sampling parameters, assuming uniform sampling in the specified range. The second block shows the parameters that are fixed in the baseline $\Lambda$CDM model with the values specified in the second column, but sampled in the minimal extensions of $\Lambda$CDM, the cosmic glitch in gravity (CGG) and dynamical dark energy (\wwCDM).
    }
    \begin{tabularx}{1.0\textwidth}{l c c X}
        \toprule
            Parameter & Fixed Value & Prior Range & Description \\
        \midrule
            $\omega_\mathrm{b} \equiv \Omega_\mathrm{b}h^2$ &      & $0.019<\omega_\mathrm{b}<0.025$         & Baryon density today.                                                          \\
            $\omega_\mathrm{c} \equiv \Omega_\mathrm{c}h^2$ &      & $0.08<\omega_\mathrm{c}<0.3$            & Cold dark matter density today.                                                \\
            $h$                                             &      & $0.4<h<0.9$                             & Hubble parameter with ${H_0 \equiv 100\,h\,\unit{\km\per\s\per\mega\parsec}}$. \\
            $\tau_\mathrm{reio}$                            &      & $0.01<\tau_\mathrm{reio}<0.2$           & Optical depth due to reionization.                                             \\
            $A_\mathrm{s}$                                  &      & $2.6 < \ln(10^{10} A_\mathrm{s}) < 3.5$ & Amplitude of the scalar primordial power spectrum.                             \\
            $n_\mathrm{s}$                                  &      & $0.9<n_\mathrm{s}<1.04$                 & Spectral index or tilt of the scalar primordial power spectrum.                \\
            $M_\nu \equiv \sum{m_\nu}$                      & \SI{0.06}{\eV} &                               & Sum of the neutrino masses, assuming a single massive neutrino.                \\
        \midrule
            $\Og$                                           & $0$  & $-0.1<\Og<0.1$                          & Cosmic glitch parameter.                                                       \\
            $w_0$                                           & $-1$ & $-2<w_0<0$                              & Constant equation-of-state parameter of dark energy.                           \\
            $w_a$                                           & 0    & $-3<w_a<2$                              & Parameter for the time-varying part of the  equation-of-state parameter of dark energy,%
                                                                                                               with additional constraint $w(a=0)={w_0+w_a<0}$ to ensure dark energy can eventually dominate.\\
        \bottomrule
    \end{tabularx}
\end{table*}

\begin{figure}[tbp]
    \centering
    \includegraphics[width=0.33\textwidth]{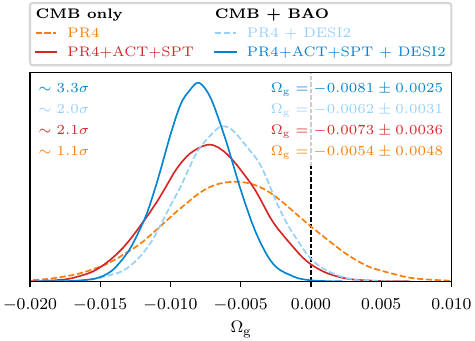}\hfill
    \includegraphics[width=0.33\textwidth]{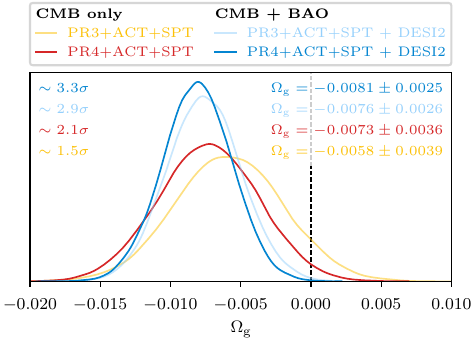}\hfill
    \includegraphics[width=0.33\textwidth]{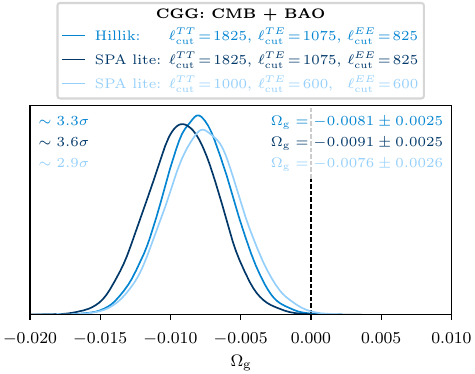}
    \caption{\label{fig:Omega_g_cmb_comp}
        The cosmic glitch parameter~\Og in light of different CMB and BAO data combinations.
        \emph{Left:}~the contribution of the high-$\ell$ ACT and SPT data on top of \Planck-PR4.
        \emph{Middle:}~the effect of the systematics treatment in the CMB likelihood, \Hillik (PR4, full and consistent foreground modeling) versus \SPA (PR3, foreground-marginalized ``lite'' likelihoods).
        \emph{Right:}~the impact of the multipole threshold $\ell_\mathrm{cut}$ used to divide the multipole range between \Planck (below $\ell_\mathrm{cut}$) and ACT (above $\ell_\mathrm{cut}$).
    }
    \includegraphics[width=0.33\textwidth]{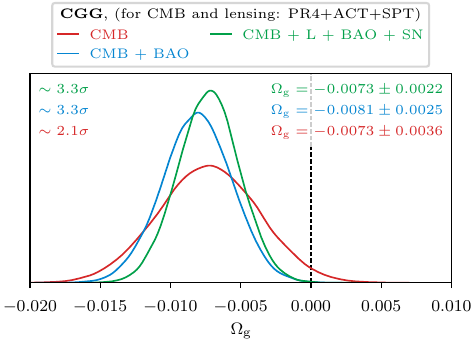}\hfill
    \includegraphics[width=0.33\textwidth]{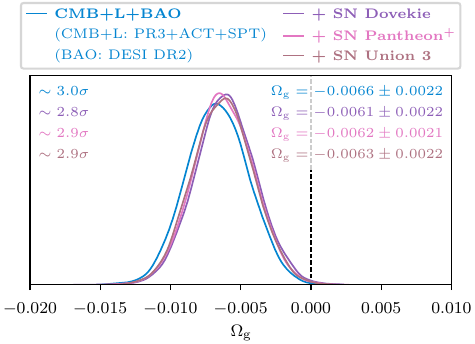}\hfill
    \includegraphics[width=0.33\textwidth]{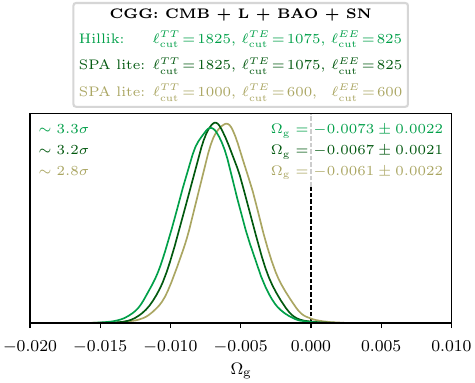}
    \caption{\label{fig:Omega_g_data_comp}
        The cosmic glitch parameter~\Og in combination with CMB lensing and SN data in addition to primary CMB and BAO data.
        \emph{Left:} the contribution of adding BAO, CMB lensing, and DES~Dovekie SN data to the primary CMB (\Hillik).
        \emph{Middle:} the negligible difference between the DES~Dovekie, Pantheon$^+$, and Union\,3 SN likelihoods (\SPA \lite for primary CMB).
        \emph{Right:} the impact of the multipole threshold $\ell_\mathrm{cut}$ (as in \cref{fig:Omega_g_cmb_comp}) on the CMB+L+BAO+SN posterior.
    }
\end{figure}

Our primary likelihood combination is CMB-Hillik plus BAO-DESI2. The CMB-Hillik combination follows the analysis in Ref.~\cite{26Tristram_Hillik}, combining \Planck-PR4, ACT~DR6, and SPT-3G~D1. To avoid double-counting between the wide-area \Planck and ACT measurements, their information is partitioned in multipole space: \Planck is used at large-to-medium angular scales and ACT at smaller scales. SPT-3G~D1 covers a smaller area within the sky observed by the other surveys and is combined under the approximation that its cross-covariance with \Planck and ACT is negligible.

\begin{itemize}
    \item \textbf{CMB-Hillik:} The CMB likelihood is the combination of two low-multipole ($\ell<30$) likelihoods, the \Planck-2018 \texttt{Commander} likelihood for low-$\ell$ $TT$~\cite{18Plancklikelihood} and the \Planck-PR4 \texttt{LoLLiPoP} likelihood for low-$\ell$ $EE$~\cite{21Tristram_Lollipop}\footnote{\url{https://github.com/planck-npipe/lollipop}}, and of three mid- to high-multipole $TTTEEE$ likelihoods combined in the \Hillik framework to use consistent foreground modeling~\cite{26Tristram_Hillik}\footnote{\url{https://github.com/mtristram/hillik}}.
    The \texttt{Hillik-Planck} likelihood corresponds essentially to the \Planck-PR4 \texttt{HiLLiPoP} likelihood~\cite{24Tristram_PR4}\footnote{\url{https://github.com/planck-npipe/hillipop}}, but is restricted here to $\ell^{TT}<1825$, $\ell^{TE}<1075$, and $\ell^{EE}<825$, so as not to overlap with the ACT~DR6 \texttt{Hillik-ACT} likelihood, which in turn is restricted to $\ell^{TT}\ge1825$, $\ell^{TE}\ge1075$, and $\ell^{EE}\ge825$. These multipole thresholds differ slightly from both the initial ACT analysis of Ref.~\cite{25ACT_DR6} and the \Hillik analysis of Ref.~\cite{26Tristram_Hillik}. We select the threshold at the cross-over point where the ACT uncertainty falls below the \Planck uncertainty (adjusted slightly so as to maximize multipole coverage where \Planck and ACT bins do not align, while enforcing strictly no overlap between the highest \Planck and lowest ACT bin).
    The \texttt{Hillik-SPT} likelihood is based on SPT-3G~D1~\cite{25Camphuis_SPT3Gd1}.
    Together, the \Hillik likelihoods have about 70 nuisance parameters, consisting of shared foreground parameters and individual calibration parameters, making the parameter-space exploration very costly, which is why we resort to the much cheaper foreground-marginalized (lite) CMB-SPA likelihood when going beyond parameter estimation to model comparison.

    \item \textbf{CMB-SPA:} For this less computationally expensive primary CMB likelihood combination, we use the same low-$\ell$ likelihoods as for CMB-Hillik (i.e., \texttt{Commander} and \texttt{LoLLiPoP}), combined with the ``lite'', i.e., foreground-marginalized, versions of the \Planck-2018~\cite{18Plancklikelihood}\footnote{\url{https://github.com/benabed/clik}}, ACT~DR6~\cite{25ACT_DR6}\footnote{\url{https://github.com/ACTCollaboration/DR6-ACT-lite}}, and SPT-3G~D1~\cite{25Camphuis_SPT3Gd1}\footnote{\url{https://github.com/SouthPoleTelescope/spt_candl_data}} likelihoods. The default multipole thresholds for splitting \Planck from ACT are $\ell_\mathrm{cut}^{TT}=1000$, $\ell_\mathrm{cut}^{TE}=600$, and $\ell_\mathrm{cut}^{EE}=600$; unless stated otherwise, we instead use the same thresholds as for CMB-Hillik. \Cref{fig:Omega_g_cmb_comp} explores the impact of this choice.

    \item \textbf{CMB-PR4:} The \Planck-PR4 combination uses \texttt{HiLLiPoP} high-$\ell$ $TTTEEE$ and \texttt{LoLLiPoP} low-$\ell$ $EE$~\cite{24Tristram_PR4}, together with \texttt{Commander} low-$\ell$ $TT$ from \Planck-2018~\cite{18Plancklikelihood}. Together with the PR4 lensing likelihood~\cite{22Carron_PR4lensing}, this corresponds to the \Planck-only combination used in previous CGG analyses~\cite{24WenCGG,24WenCGG_essay}.

    \item \textbf{L-SPA:} When adding CMB lensing, we use the \Planck+ACT+SPT lensing likelihood from Ref.~\cite{25Ge_SPT3GMUSE}\footnote{\url{https://github.com/qujia7/spt_act_likelihood}}, based on the \Planck-PR4 and ACT~DR6 lensing analyses~\cite{22Carron_PR4lensing,24Madhavacheril_ACTDR6lensing,24Qu_ACTDR6lensing}.

    \item \textbf{BAO-DESI2:} We use the BAO measurements from DESI~DR2, the first three years of DESI observations~\cite{25DESI_DR2BAO}, through the public \texttt{desi\_bao\_all} likelihood in \texttt{Cobaya}.\footnote{\url{https://github.com/CobayaSampler/cobaya/blob/master/cobaya/likelihoods/bao/desi_2024_bao_all.yaml}}

    \item \textbf{Supernovae:} We add DES Dovekie~\cite{25Dovekie_DESY5SNe}\footnote{\url{https://github.com/CobayaSampler/cobaya/blob/master/cobaya/likelihoods/sn/desdovekie.yaml}}, Pantheon$^+$~\cite{21Scolnic_PantheonP}\footnote{\url{https://github.com/CobayaSampler/cobaya/blob/master/cobaya/likelihoods/sn/pantheonplus.yaml}}, and Union\,3~\cite{25Rubin_UnionY3}\footnote{\url{https://github.com/CobayaSampler/cobaya/blob/master/cobaya/likelihoods/sn/union3.yaml}} one at a time. For a comparison of their impact, see \cref{fig:Omega_g_data_comp}.

\end{itemize}

Because a single \Hillik chain is roughly one to two orders of magnitude more expensive than an \SPA \lite chain, we use \Hillik for the headline parameter constraints (\cref{eq:main_result}, \cref{tab:constraints}) and \SPA \lite for the more expensive nested-sampling model comparison. The two likelihoods give consistent results wherever both are available. For CMB+BAO, \Hillik gives $\Og=-0.0081\pm0.0025$, and \SPA \lite gives $-0.0076\pm0.0026$ for the \SPA \lite default multipole cut and $-0.0091\pm0.0025$ for a multipole cut matching the one of \Hillik; for the primary CMB alone, $-0.0073\pm0.0036$ versus $-0.0058\pm0.0039$ versus $-0.0077\pm0.0037$; and for CMB+L+BAO+SN, $-0.0073\pm0.0022$ versus $-0.0061\pm0.0022$ versus $-0.0067\pm0.0021$ (\cref{fig:Omega_g_cmb_comp,fig:Omega_g_data_comp,fig:triangles_cgg_hillik_vs_spa}).
\SPA \lite with the low multipole cut (${\ell_\mathrm{cut}^{TT}=1000}$, ${\ell_\mathrm{cut}^{TE}=600}$, and ${\ell_\mathrm{cut}^{EE}=600}$) thus returns slightly less negative central values than \Hillik, by $0.1$ to $0.4\,\sigma$.
\SPA \lite with the high multipole cut (${\ell_\mathrm{cut}^{TT}=1825}$, ${\ell_\mathrm{cut}^{TE}=1075}$, and ${\ell_\mathrm{cut}^{EE}=825}$), on the other hand, returns slightly more negative central values than \Hillik, by $0.1$ to $0.3\,\sigma$ (with the exception of also adding CMB lensing and SN data).

\subsection{Constraints and data progression}
\label{sec:sm_constraints}

BAO measurements constrain the transverse and radial distances, $D_\mathrm{M}/r_\mathrm{d}$ and $D_\mathrm{H}/r_\mathrm{d}$, corresponding to the two Alcock--Paczynski degrees of freedom, which can be compressed into the angle-averaged distance $D_\mathrm{V}/r_\mathrm{d}$; for the BGS tracer only the isotropic $D_\mathrm{V}/r_\mathrm{d}$ is available. These are the quantities compared with the CMB-only model predictions in \cref{fig:bao_residuals} of the main text. \Cref{tab:constraints} lists the marginalized constraints on all sampled and the main derived parameters for \LCDM, CGG, and \wwCDM, for CMB+BAO and for CMB+L+BAO+SN. Relative to \LCDM, CGG shifts $H_0$ up by approximately ${\SI{0.1}{\km\per\s\per\mega\parsec}}$, \Om down by ${0.003}$, and the sound horizon $r_\mathrm{d}$ up by ${\SI{0.6}{\percent}}$, so that $r_\mathrm{d}h$ moves toward the value preferred by DESI. It also raises $\sigma_8$ and $S_8$ by about \SIrange[range-units=single,range-phrase=--]{2}{3}{\percent}, in line with the faster growth of \cref{eq:sm_growth} above. \wwCDM on CMB+BAO alone instead drives $H_0$ down to around ${\SI{64}{\km\per\s\per\mega\parsec}}$ with a very broad posterior; only the addition of SN data pulls it back to ${\SI{67}{\km\per\s\per\mega\parsec}}$.

\begin{table}[tbp]
    \centering
    \small
    \setlength{\tabcolsep}{9pt}
    \caption{\label{tab:sm_summary}
        Constraints on \Og across data combinations. The significance is $|\langle\Og\rangle|/\sigma_{\Og}$ for the 1-dimensional marginalized posterior (except for the very skewed posterior of the $\log_{10}\cs^2$ marginalized run, for which we quote the $1\,\sigma$ equivalent iso-probability bounds instead of the standard deviation, and significance is estimated from the iso-probability level at $\Og=0$ and expressed in a Gaussian-equivalent sigma value). See also \cref{fig:Omega_g,fig:cs2_2d}, as well as \cref{fig:Omega_g_cmb_comp,fig:Omega_g_data_comp}.
    }
    \begin{tabular}{lcc}
    \toprule
        Data combination & \Og & Significance \\
    \midrule
        \multicolumn{3}{l}{\itshape Previous analyses}\\
        \quad CMB \Planck-2018 + L \Planck-2018 \cite{24WenCGG}           & $-0.0087 \pm 0.0046$ & $1.9\,\sigma$ \\
        \quad CMB-PR4 + L-PR4 \cite{24WenCGG}                  & $-0.0054 \pm 0.0042$ & $1.3\,\sigma$ \\
        \quad CMB-PR4 + L-PR4 + BAO-DESI1 \cite{24WenCGG_essay} & $-0.0067 \pm 0.0029$ & $2.3\,\sigma$ \\
    \midrule
        \multicolumn{3}{l}{\itshape This work} \\
        \quad CMB-PR4 + BAO-DESI2                       & $-0.0062 \pm 0.0031$ & $2.0\,\sigma$ \\
        \quad CMB-SPA + BAO-DESI2 (with $\ell_\mathrm{cut}=1000, \hphantom{1}600, 600$) & $-0.0076 \pm 0.0026$ & $2.9\,\sigma$ \\
        \quad CMB-SPA + BAO-DESI2 (with $\ell_\mathrm{cut}=1825, 1075, 825$) & $-0.0091 \pm 0.0025$ & $3.6\,\sigma$ \\
        \quad \textbf{CMB-Hillik $\bm{+}$ BAO-DESI2} & $\boldsymbol{-0.0081 \pm 0.0025}$ & $\boldsymbol{3.3\,\sigma}$ \\
    \midrule
        \multicolumn{3}{l}{\itshape Extension I: CMB lensing and supernovae}\\
        \quad CMB-SPA + L-SPA + BAO-DESI2 & $-0.0066 \pm 0.0022$ & $3.0\,\sigma$ \\
        \quad CMB-SPA + L-SPA + BAO-DESI2 + SN-Pantheon$^+$ & $-0.0062 \pm 0.0021$ & $2.9\,\sigma$ \\
        \quad CMB-SPA + L-SPA + BAO-DESI2 + SN-Union3 & $-0.0063 \pm 0.0022$ & $2.9\,\sigma$ \\
        \quad CMB-SPA + L-SPA + BAO-DESI2 + SN-Dovekie (with $\ell_\mathrm{cut}=1000, \hphantom{1}600, 600$) & $-0.0061 \pm 0.0022$ & $2.8\,\sigma$ \\
        \quad CMB-SPA + L-SPA + BAO-DESI2 + SN-Dovekie (with $\ell_\mathrm{cut}=1825, 1075, 825$) & $-0.0067 \pm 0.0021$ & $3.2\,\sigma$ \\
        \quad \textbf{CMB-Hillik $\bm{+}$ L-SPA $\bm{+}$ BAO-DESI2 $\bm{+}$ SN-Dovekie} & $\boldsymbol{-0.0073 \pm 0.0022}$ & $\boldsymbol{3.3\,\sigma}$ \\
    \midrule
        \multicolumn{3}{l}{\itshape Extension II: CMB-Hillik + BAO-DESI2}\\
        \quad \dots with $\cs^2=1$ & $-0.0081 \pm 0.0025$ & $3.3\,\sigma$ \\
        \quad \dots with $\cs^2=1000$ & $-0.0059 \pm 0.0020$ & $2.9\,\sigma$ \\
        \quad \dots with $\cs^2=0.001$ & $-0.0216 \pm 0.0069$ & $3.1\,\sigma$ \\
        \quad \dots with $\log_{10}\cs^2$ marginalized over $[-5,5]$ & $-0.0147_{~-~0.0035~\,}^{~+~0.0079~\,}$ & $2.8\,\sigma$ \\
    \bottomrule
    \end{tabular}
\end{table}

\begin{figure}
    \centering
    \includegraphics[width=0.49\textwidth]{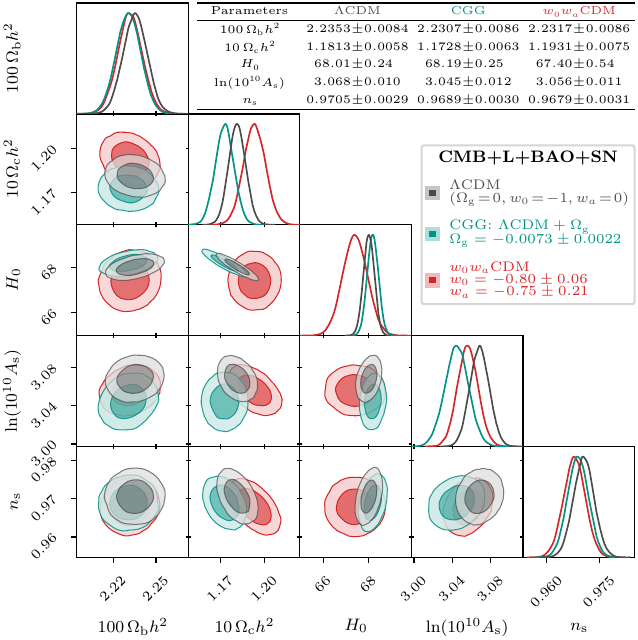}\hfill
    \includegraphics[width=0.49\textwidth]{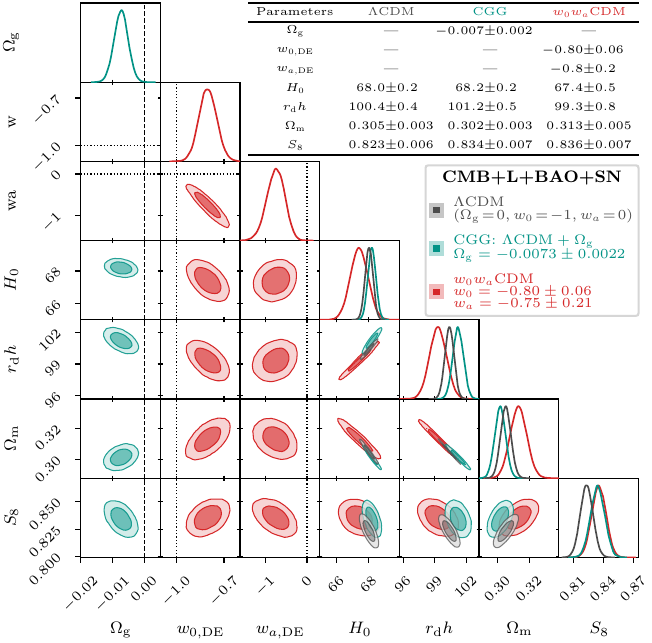}
    \caption{
        Model comparison between \LCDM, CGG (i.e., $\LCDM+\Og$), and \wwCDM on CMB+L+BAO+SN data.
        The left triangle shows the cosmological sampling parameters (except $\tau_\mathrm{reio}$); the right triangle shows derived parameters of interest for cosmological tensions.
    }
    \label{fig:triangles_model_comp}
\end{figure}

\begin{figure}[tbp]
    \centering
    \includegraphics[width=0.385\textwidth]{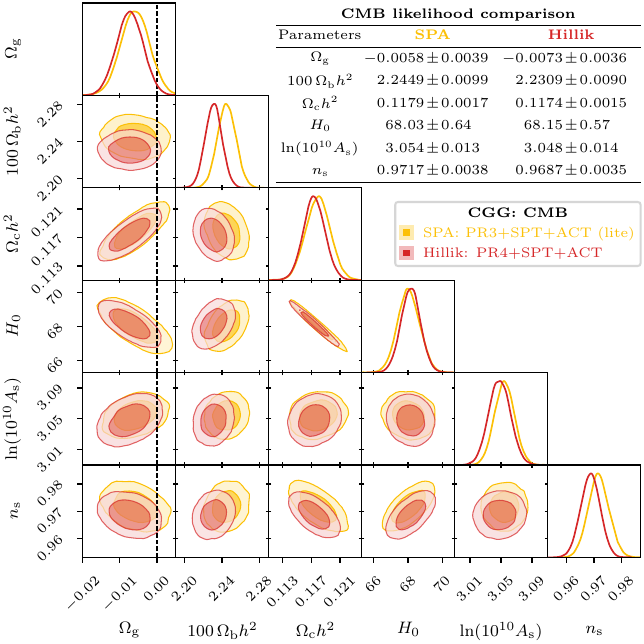}\qquad
    \includegraphics[width=0.385\textwidth]{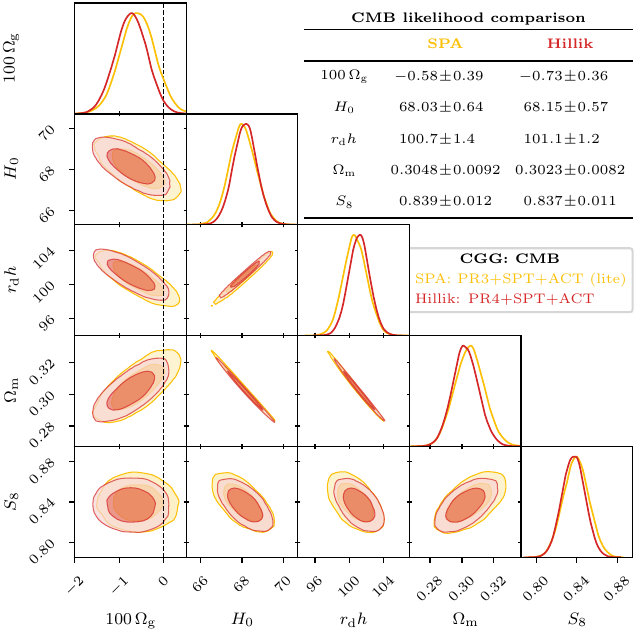}\\\medskip
    \includegraphics[width=0.385\textwidth]{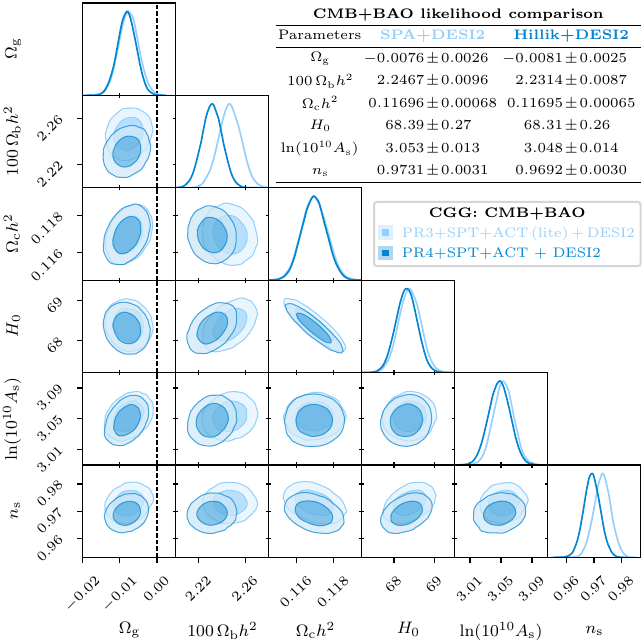}\qquad
    \includegraphics[width=0.385\textwidth]{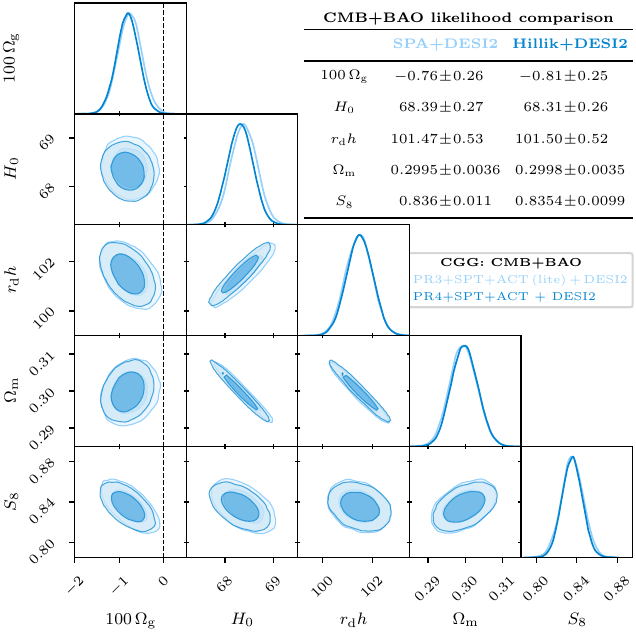}\\\medskip
    \includegraphics[width=0.385\textwidth]{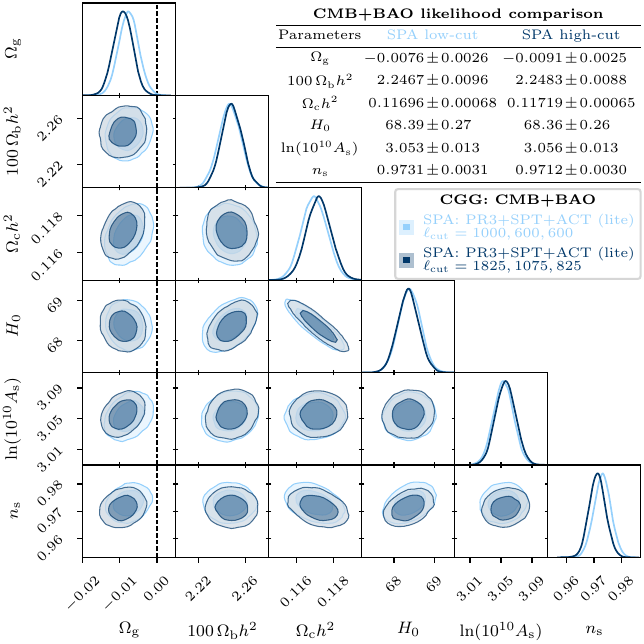}\qquad
    \includegraphics[width=0.385\textwidth]{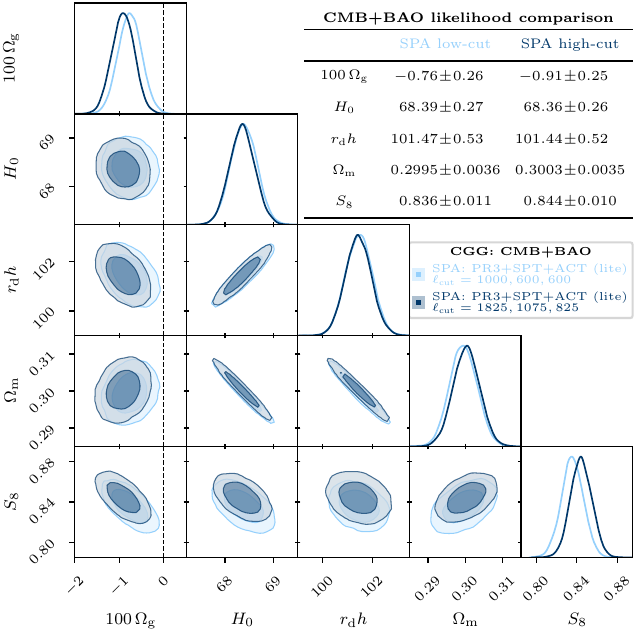}
    \caption{
        Comparison of CGG posteriors under the \SPA \lite and \Hillik CMB likelihoods.
        The left triangles show the cosmological sampling parameters (except the optical depth~$\tau_\mathrm{reio}$); the right show derived parameters of interest for cosmological tensions.
        The top two triangles use only CMB data; the middle two combine the same CMB data with DESI~DR2 BAO; the bottom two compare the influence of the \Planck/ACT multipole thresholds~$\ell_\mathrm{cut}$ ($1000, 600, 600$ for a low cut or $1825, 1075, 825$ for a high cut for $TT$, $TE$, and $EE$, respectively).}
    \label{fig:triangles_cgg_hillik_vs_spa}
\end{figure}

The progression in \cref{fig:Omega_g_cmb_comp,fig:Omega_g_data_comp} and \cref{tab:sm_summary} shows that the negative-\Og preference is not introduced by a single new data set. It is already present in the earlier \Planck-PR4 + DESI~DR1 analysis, remains after updating the BAO to DESI~DR2, and becomes stronger in the full CMB combination of \Planck+ACT+SPT, both with the \Hillik and with the \SPA \lite likelihoods. The choice of \Planck/ACT multipole threshold moves the CMB+L+BAO+SN constraint between $-0.0061\pm0.0022$ and $-0.0067\pm0.0021$ (right panel of \cref{fig:Omega_g_data_comp}), i.e., well within $1\,\sigma$. \Cref{fig:triangles_cgg_hillik_vs_spa} compares the full CGG posteriors under the \Hillik and \SPA likelihoods, and \cref{fig:triangles_model_comp} compares the three models on CMB+L+BAO+SN.

\subsection{Supernova robustness}
\label{sec:sm_sn}

Supernova data play a prominent role in current claims for dynamical dark energy in the $w_0w_a$ parameterization. For CGG, their impact is small. Adding Pantheon$^+$, Union\,3, or DES Dovekie shifts \Og only mildly toward zero while leaving the preference near $3\,\sigma$ (\cref{fig:Omega_g_data_comp} and \cref{tab:sm_summary}). All SN data sets prefer a slightly larger matter density \Om than both BAO and CMB (for \LCDM and CGG). Since more negative \Og is correlated with lower \Om and higher $H_0$, pulling \Om upward also moves \Og toward zero.

\subsection{Bayesian model comparison}
\label{sec:sm_bmc}

For a Bayesian model comparison, we resorted to the \SPA \lite likelihood, because the full \Hillik likelihood comes with around $70$ nuisance parameters, making it too computationally expensive for calculating both the best-fit and nested-sampling analyses with sufficiently small sampling uncertainties for conclusive statements.

\begin{figure}[t]
    \centering
    \includegraphics[width=0.6\textwidth]{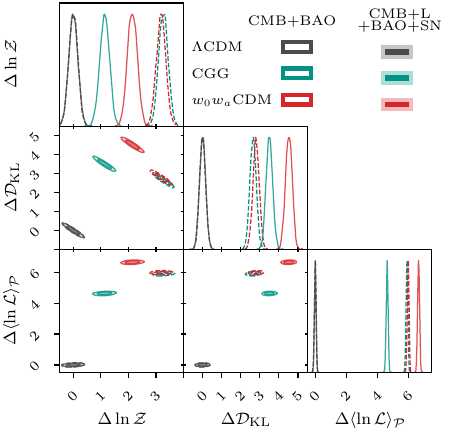}
    \caption{
        Evidence decomposition for \LCDM, CGG, and \wwCDM on CMB+BAO (unfilled, dashed) and on CMB+L+BAO+SN (filled, solid) data, normalized to \LCDM. The log-evidence $\ln\mathcal{Z}$ is decomposed into the posterior-averaged log-likelihood $\langle\ln\mathcal{L}\rangle_\mathcal{P}$ (goodness of fit) and the Kullback--Leibler divergence $\mathcal{D}_\mathrm{KL}$ between posterior and prior (Occam penalty), $\ln\mathcal{Z}=\langle\ln\mathcal{L}\rangle_\mathcal{P}-\mathcal{D}_\mathrm{KL}$; see Ref.~\cite{26Hergt_BMCtension} for a discussion of these statistics.
        The $\Delta$ indicates normalization with respect to \LCDM.
        Because of the high computational cost of nested sampling in high dimensions, these runs use the \SPA \lite likelihood for the primary CMB rather than \Hillik, but with the same \Planck/ACT multipole thresholds ($\ell_\mathrm{cut}^{TT}=1825$, $\ell_\mathrm{cut}^{TE}=1075$, and $\ell_\mathrm{cut}^{EE}=825$).
    }
    \label{fig:model_comp_triangle}
\end{figure}

We compare CGG, \LCDM, and \wwCDM using the evidence ratio relative to \LCDM, $\Delta\ln\mathcal{Z}_X=\ln\mathcal{Z}_X-\ln\mathcal{Z}_{\LCDM}$ (\cref{fig:model_comp_triangle}). For CMB+BAO (\SPA \lite, with $\ell_\mathrm{cut}^{TT}=1825$, $\ell_\mathrm{cut}^{TE}=1075$, and $\ell_\mathrm{cut}^{EE}=825$ as in \Hillik), both extensions are moderately favored over \LCDM, with $\Delta\ln\mathcal{Z}=+3.2\pm0.2$ for CGG and $+3.2\pm0.2$ for \wwCDM: \wwCDM fits slightly better ($\Delta\langle\ln\mathcal{L}\rangle_{\mathcal P}$ larger by about one unit) but pays a correspondingly larger Occam penalty for its second parameter. When CMB lensing and SN data are added, the preference for CGG over \LCDM weakens to $\Delta\ln\mathcal{Z}=+1.1\pm0.2$, since the SN data pull \Om upward and hence \Og toward zero, whereas \wwCDM is still slightly favored by $+2.2\pm0.2$ with the \SPA \lite, because its extra freedom can accommodate the mismatch in \Om between SNe and BAOs.

Evidence ratios are prior dependent and should be interpreted at fixed prior volume. Since the \Og prior is much wider than the posterior, changing its width $W$ shifts $\Delta\ln\mathcal{Z}_\mathrm{CGG}$ by approximately $-\ln(W/W_\mathrm{ref})$, with $W_\mathrm{ref}=0.2$ corresponding to the sampled range $[-0.1,0.1]$ in \cref{tab:priors}: halving the prior width gains $\ln 2\simeq0.7$, and a prior 4 times narrower than our default would move CGG from a moderate to strong preference using CMB+BAO. For \wwCDM, the condition $w_0+w_a<0$ imposed to preserve matter domination reduces the effective prior volume relative to the rectangular prior of \cref{tab:priors}, which slightly softens its Occam penalty. Conversely, an extension parameter that the data leave unconstrained, as is the case for $w_a$ (and largely $w_0$) without SN data, incurs almost no penalty regardless of prior width, so that the evidence does not reward the predictive advantage of the 1-parameter model seen in \cref{fig:bao_residuals} of the main text. For a flat prior that fully contains the posterior this rescaling is exact, so no additional nested-sampling runs are needed to translate our numbers to another prior width.

\end{document}